\documentclass[prd,twocolumn,showpacs,nofootinbib,floats,floatfix]{revtex4}

\usepackage{graphicx}
\usepackage{amsmath,amsfonts}
\usepackage{dcolumn}
\usepackage{color}

\newcommand{\td}[1]{{\rm d}#1} 
\newcommand{\Snormal}{n} 
\newcommand{\Lapse}{\alpha} 
\newcommand{\CLapse}{\tilde\alpha} 
\newcommand{\Shift}{\beta} 
\newcommand{\SMetric}{\gamma} 
\newcommand{\CMetric}{\tilde\gamma} 
\newcommand{\CF}{\psi} 
\newcommand{\ExCurv}{K} 
\newcommand{\TrExCurv}{K} 
\newcommand{\TFExCurv}{A} 
\newcommand{\CTFExCurv}{\tilde{A}} 
\newcommand{\SRicci}{{R}} 
\newcommand{\SRicciS}{{R}} 
\newcommand{\CRicciS}{\tilde{R}} 
\newcommand{\dtime}{\partial_t} 
\newcommand{\SCD}{{\nabla\!}} 
\newcommand{\CCD}{{\tilde\nabla}\!} 
\newcommand{\LieD}[1]{{{\cal L}_{#1}}}
\newcommand{\CLD}[1]{(\tilde{\mathbb L}{#1})} 

\newcommand{\CMtd}{\tilde{u}} 
\newcommand{\Bnormal}{s} 
\newcommand{\CBnormal}{\tilde{s}} 
\newcommand{\BMetric}{h} 
\newcommand{\CBMetric}{\tilde{h}} 
\newcommand{\ONull}{k} 
\newcommand{\Oexpansion}{\theta} 
\newcommand{\Oshear}{\sigma} 
\newcommand{\BCD}{D} 
\newcommand{\CBCD}{\tilde{D}} 
\newcommand{\BRicci}{{{}^2\!R}} 
\newcommand{\BRicciS}{{{}^2\!R}} 
\newcommand{\perpShift}{{\Shift_{\!\mbox{\tiny$\perp$}}}} 
\newcommand{\parShift}{{\Shift_{\mbox{\tiny$\parallel$}}}} 
\newcommand{\SLap}{{{}_{\mbox{\tiny s}}\!\!\nabla^2}} 
\newcommand{\SGrad}{{{}_{\mbox{\tiny s}}\!\!\vec{\nabla}}} 
\newcommand{\SDiv}{{{}_{\mbox{\tiny s}}\!\!\vec{\nabla}\cdot}} 
\newcommand{\SVort}{{{}_{\mbox{\tiny s}}\!\!\vec{\nabla}\times}} 

\begin{document}

\title{Circular orbits and spin in black-hole initial data}

\author{Matthew Caudill}\email{mscaudil@artsci.wustl.edu}
\altaffiliation[\\Current address:]{
Department of Physics, Washington University in St.\ Louis, 
St.\ Louis, Missouri\ \ 63130}
\author{Gregory B. Cook}\email{cookgb@wfu.edu}
\author{Jason D. Grigsby}\email{grigjd3@wfu.edu}
\affiliation{Department of Physics, Wake Forest University,
		 Winston-Salem, North Carolina\ \ 27109}
\author{Harald P. Pfeiffer}\email{harald@tapir.caltech.edu}
\affiliation{Theoretical Astrophysics, California Institute of Technology,
 		 Pasadena, California\ \ 91125}

\date{\today}

\begin{abstract}
The construction of initial data for black-hole binaries usually
involves the choice of free parameters that define the spins of the
black holes and essentially the eccentricity of the orbit.  Such
parameters must be chosen carefully to yield initial data with the
desired physical properties. In this paper, we examine these choices
in detail for the quasiequilibrium method coupled to
apparent-horizon/quasiequilibrium boundary conditions.  First, we
compare two independent criteria for choosing the orbital frequency,
the ``Komar-mass condition'' and the ``effective-potential method,''
and find excellent agreement.  Second, we implement quasi-local
measures of the spin of the individual holes, calibrate these with
corotating binaries, and revisit the construction of {\em
non-spinning} black hole binaries.  Higher-order effects, beyond those
considered in earlier work, turn out to be important.  Without those,
supposedly non-spinning black holes have appreciable quasi-local spin;
furthermore, the Komar-mass condition and effective potential method agree
only when these higher-order effects are taken into account.  We
compute a new sequence of quasi-circular orbits for non-spinning
black-hole binaries, and determine the innermost stable circular orbit
of this sequence.
\end{abstract}


\maketitle

\section{Introduction}
\label{sec:introduction}

Recently, significant progress has been made in numerically evolving
black-hole
binaries\cite{Pretorius-2005,Campanelli-etal-2006a,Baker-etal-2006a,Diener-etal-2005}.
A major goal of these simulations is to estimate the gravitational
waveform produced by astrophysical black-hole binaries.  These
waveforms will ultimately be used to aid in the detection and
interpretation of the gravitational wave signals we expect to see in
observatories such as LIGO, VIRGO, TAMA, and GEO600.  In order for such
simulations to yield astrophysically relevant results, the initial
data must be constructed to be astrophysically realistic.

A very effective approach for constructing numerical black-hole binary
initial data has been developed and explored by two of the
authors\cite{cook-pfeiffer-2004a} (see also
Refs.~\cite{gourgoulhon-etal-2002a,gourgoulhon-etal-2002b,Cook-2002}).
In Ref.~\cite{cook-pfeiffer-2004a}, the authors focused attention on
two specific, limiting cases of binary initial data: corotating black
holes and irrotational (non-spinning) black holes.  These numerical
initial-data solutions were compared against previous numerical
results\cite{cook94e,gourgoulhon-etal-2002b} and to analytic
post-Newtonian estimates\cite{Damour-etal-2002,Blanchet:2002} for
binaries in circular orbits and appear to give the best agreement yet
between numerical and analytic models of close black-hole binaries in
circular orbits.  However, various aspects of the physical content of
these initial-data sets have not been fully tested.

In order to construct quasi-circular orbits (as opposed to general
elliptical orbits), Gourgoulhon et~al.\cite{gourgoulhon-etal-2002a} proposed that the data must satisfy a
simple condition: the Komar\cite{komar59} and the ADM\cite{ADM} masses
must agree if the orbits are quasi-circular.  The Komar mass is only a
reasonable definition of the total mass of a system if the system is
{\em stationary}.  The ADM mass, is an invariant measure of the total
mass of a system as measured at space-like infinity.  So, for binary
systems that are quasi-stationary when they are in quasi-circular
orbit, the ansatz seems quite reasonable.  This ansatz has been tested
by comparing numerical models with analytic post-Newtonian models.  It
has also been tested in the case of neutron-star
binaries\cite{taniguchi-gourgoulhon-2002} where an independent method
of determining circular orbits exists.  For black-hole binaries, an
independent method of determining circular orbits exists in the
so-called {\em effective potential method}\cite{cook94e}.  In this
paper, we will further test the Komar-ADM mass ansatz by constructing
circular orbits using the effective-potential method.

Another important aspect of the binary initial data constructed in
Ref.~\cite{cook-pfeiffer-2004a} that has not been adequately verified
relates to the spins of the individual black holes.  In
Ref.~\cite{cook-pfeiffer-2004a}, the spin of each black hole is 
fixed by a particular choice of boundary conditions applied at the
surface of the black hole.  For the case of corotating black holes,
the choice of boundary conditions is unambiguous.  However, for
non-spinning black holes, the boundary conditions were chosen
in a way that should be correct in the limit of large separation
between the black holes in the binary.  However, this non-spinning
ansatz has not yet been checked.

Of course, we must keep in mind that the angular momentum (spin) of an
individual black hole in a close binary system is not rigorously
defined in general relativity.  To measure the spin, we will have to
rely on a quasi-local definition.  There are many such
definitions\cite{Szabados-2004}.  For our purposes, we will use a
definition first made rigorous by Brown and
York\cite{brown-york-1993} and also derived within the more recent {\em
isolated horizons} framework of Ashtekar and
Krishnan\cite{ashtekar-krishnan-2003a,Ashtekar-Krishnan-2004}.
We will explore extensively the spins of the individual black holes
in our binary initial data using this quasi-local definition of the
angular momentum.  A major result of this study is that we must
refine our method for setting the boundary conditions in order to
construct models where the black holes are not rotating.

We begin in Sec.~\ref{sec:init-data-form} with a review of the entire
formalism used to construct our initial data.  In
Sec.~\ref{sec:comp-quasi-local} we discuss various issues associated
with computing the quasi-local spin of a black hole. In
Sec.~\ref{sec:corotating-binaries} we examine the case of corotating
binaries, focusing first on exploring the criteria for defining
circular orbits and then examining the spins of corotating black
holes.  In Sec.~\ref{sec:non-spin-binary} we turn to the case of
non-spinning binaries and define the correct approach for obtaining
non-spinning black holes.  Finally, in Sec.~\ref{sec:discussion} we
present results related to the inner-most stable circular orbits for
both non-spinning and corotating binaries, and review the major points
from the paper.

\section{Initial-data Formalism}
\label{sec:init-data-form}

The black-hole initial-data sets that we consider below are
constructed using the conformal thin-sandwich
decomposition\cite{york-1999,Pfeiffer-York-2003}, a set of boundary
conditions imposed on the black-hole excision
surfaces\cite{Cook-2002,cook-pfeiffer-2004a} and at asymptotic
infinity, and a set of assumptions for various freely-specifiable
fields.  Below, we will outline the most important details of the
various pieces of our approach.

\subsection{The conformal thin-sandwich decomposition}
\label{sec:conf-thin-sandw}

In this work, we will use the standard \mbox{3+1} decomposition with the interval written as
\begin{equation}
\td{s}^2 = -\Lapse^2 \td{t}^2 
  + \SMetric_{ij}(\td{x}^i +\Shift^i\td{t})(\td{x}^j +\Shift^j\td{t}),
\label{eq:3+1_interval}
\end{equation}
where $\SMetric_{ij}$ is the 3-metric induced on a $t=\mbox{const.}$
spatial hypersurface, $\Lapse$ is the lapse function, and $\Shift^i$
is the shift vector.  The extrinsic curvature of the spatial slice,
$\ExCurv_{ij}$, is defined by
\begin{equation}
\ExCurv_{\mu\nu} \equiv -\frac12\SMetric^\delta_\mu\SMetric^\rho_\nu
          \LieD{\Snormal}\SMetric_{\delta\rho},
\label{eq:ExCurv_def}
\end{equation}
where $\LieD{\Snormal}$ denotes the Lie derivative along the unit
normal to the spatial slice, $\Snormal^\mu$.  Einstein's equations, in
vacuum, then reduce to four sets of equations.  Two are evolution
equations for the spatial metric and extrinsic curvature:
\begin{equation}
\dtime\SMetric_{ij} = -2\Lapse\ExCurv_{ij} + 2\SCD_{(i}\Shift_{j)},
\label{eq:metric_evol}
\end{equation}
and
\begin{eqnarray}
\dtime\ExCurv_{ij} &=& -\SCD_i\SCD_j\Lapse 
+ \Lapse\left[\SRicci_{ij} 
- 2\ExCurv_{i\ell}\ExCurv^\ell_j 
+ \TrExCurv\ExCurv_{ij}\right]
\nonumber \\ && \mbox{}
+\Shift^\ell\SCD_\ell\ExCurv_{ij} 
+ 2\ExCurv_{\ell(i}\SCD_{j)}\Shift^\ell.
\label{eq:ExCurv_evol}
\end{eqnarray}
The remaining two are the constraint equations
\begin{equation}
\SRicciS + \TrExCurv^2 - \ExCurv_{ij}\ExCurv^{ij} = 0
\label{eq:Hamiltonian_const}
\end{equation}
and
\begin{equation}
\SCD_j(\ExCurv^{ij} - \SMetric^{ij}\TrExCurv) = 0.
\label{eq:Momentum_const}
\end{equation}
Here, $\SCD_i$, $\SRicci_{ij}$, and $\SRicciS$ are, respectively, the
covariant derivative, Ricci tensor, and Ricci scalar associated with
the spatial metric $\SMetric_{ij}$.  Finally, the trace of the extrinsic
curvature is denoted $\TrExCurv \equiv \ExCurv^i_i$.

The conformal thin-sandwich decomposition employs a York--Lichnerowicz
conformal decomposition of the metric and various other
quantities\cite{lichnerowicz-1944,york-1971,york-1972}.  The conformal
factor, $\CF$, is defined via
\begin{equation}
\SMetric_{ij} \equiv \CF^4\CMetric_{ij},
\label{eq:Metric_decomp}
\end{equation}
where $\CMetric_{ij}$ is a ``conformal metric''.  The time derivative
of the conformal metric is introduced by the definitions
\begin{align}
\CMtd_{ij} &\equiv \dtime\CMetric_{ij} \label{eq:Conf_u_def},\\
\CMetric^{ij}\CMtd_{ij} &\equiv 0. \label{eq:Tr_u_def}
\end{align}
From this, it follows that the tracefree extrinsic curvature
$\TFExCurv^{ij}\equiv\TrExCurv^{ij}-\frac13\SMetric^{ij}\TrExCurv$ takes
the form
\begin{equation}
\TFExCurv^{ij} =\frac{\CF^{-10}}{2\CLapse}\left[\CLD{\Shift}^{ij} -
\CMtd^{ij}\right],
\label{eq:ExCurv_decomp}
\end{equation}
where $\CLapse\equiv\CF^{-6}\Lapse$ is the conformal lapse function, and
$\CMtd^{ij}=\CMtd_{kl}\CMetric^{ik}\CMetric^{jl}$. 
Furthermore, $\CLD{V}$ is the conformal-Killing (or longitudinal)
operator acting on a vector, defined by
\begin{equation}
\CLD{V}_{ij} \equiv 2\CCD_{(i}V_{j)} - \mbox{$\frac23$}\CMetric_{ij}\CCD_kV^k,
\label{eq:conf-Killing_op}
\end{equation}
where $\CCD_k$ is the covariant derivative compatible with
$\CMetric_{ij}$.  Finally, the conformal tracefree extrinsic
curvature can be written as
\begin{equation}
\CTFExCurv^{ij} \equiv \CF^{10}\TFExCurv^{ij}= \frac{1}{2\CLapse}
\left[\CLD{\Shift}^{ij} - \CMtd^{ij}\right].
\label{eq:CTFExcurv_def}
\end{equation}

In terms of our conformally decomposed variables, the Hamiltonian
constraint (\ref{eq:Hamiltonian_const}) can be written
\begin{equation}
\CCD\,^2\CF - \mbox{$\frac18$}\CF\CRicciS
- \mbox{$\frac1{12}$}\CF^5\TrExCurv^2
+ \mbox{$\frac18$}\CF^{-7}\CTFExCurv_{ij}\CTFExCurv^{ij} = 0,
\label{eq:CTS_Ham_con}
\end{equation}
where $\CRicciS$ is the Ricci scalar associated with $\CMetric_{ij}$,
and the momentum constraint (\ref{eq:Momentum_const}) as
\begin{equation}
\CCD_j\left(\mbox{$\frac{1}{2\CLapse}$}\CLD\Shift^{ij}\right)
-\mbox{$\frac{2}{3}$}\CF^6\CCD^{\;i}\TrExCurv 
       - \CCD_j\left(\mbox{$\frac1{2\CLapse}$}\CMtd^{ij}\right)=0.
\label{eq:CTS_mom_con}
\end{equation}
Furthermore, the trace of Eq.~(\ref{eq:ExCurv_evol}) can
be written as
\begin{eqnarray}
\CCD\,^2(\CF^7\CLapse)
-(\CF^7\CLapse)\left( \mbox{$\frac18$}\CRicciS
+ \mbox{$\frac5{12}$}\CF^{4}\TrExCurv^2 
+ \mbox{$\frac78$}\CF^{-8}\CTFExCurv_{ij}\CTFExCurv^{ij}\right)
\nonumber \\ \mbox{}
=- \CF^5\left(\dtime\TrExCurv-\Shift^k\CCD_k\TrExCurv\right).
\label{eq:Const_TrK_eqn}
\end{eqnarray}

Within the conformal thin-sandwich formalism, the fundamental
variables are: $\CF$, $\CLapse$, $\Shift^i$, $\CMetric_{ij}$,
$\TrExCurv$, $\dtime\CMetric_{ij}\equiv\CMtd_{ij}$, and
$\dtime\TrExCurv$.  Of these, $\CMetric_{ij}$, $\TrExCurv$,
$\CMtd_{ij}$, and $\dtime\TrExCurv$ represent the eight gauge and
dynamical degrees of freedom of the gravitational field.  These fields
must be chosen based on the physics of the initial data one wishes to
model.  The remaining fields, $\CF$, $\CLapse$, and $\Shift^i$
represent the constrained degrees of freedom.  Once the other fields
have been fixed, these fields are determined by solving
Eqs.~(\ref{eq:CTS_Ham_con}--\ref{eq:Const_TrK_eqn}) as a set of
coupled elliptic equations.

Formulating a well-posed elliptic system requires that
we impose boundary conditions.  Typically, these systems are solved
under the assumption that the spacetime is asymptotically flat.  If we
let $r$ denote a coordinate radius measured from the location of the
center of energy of the system, then as $r\to\infty$ we have that
\begin{subequations}\label{eq:outer-BCs}
\begin{align}
\label{eq:CF_BC_infty}
\CF\big|_{r\to\infty}&=1,\\
\label{eq:shift_BC_infty}
\Shift^i\big|_{r\to\infty}&=({\mathbf\Omega_0}\times{\bf r})^i,\\
\label{eq:Lapse_BC_infty}
\Lapse\big|_{r\to\infty}=\CLapse\big|_{r\to\infty}&=1.
\end{align}
\end{subequations}
$\Omega_0$ is the orbital angular velocity of a binary system, or the
rotational angular velocity of a single object, as measured at
infinity.  The boundary condition on the shift is chosen so that the
time coordinate, $t^\mu = \Lapse\Snormal^\mu + \Shift^\mu$, is helical
and tracks the rotation of the
system\cite{cook96b,baumgarte_etal98b,gourgoulhon-etal-2002a,Cook-2002,cook-pfeiffer-2004a}.
If we wish to consider systems with one or more black holes, and if we
excise the interior of the black hole to avoid difficulties with
singularities, then we must also impose boundary conditions on the
excision surfaces.

\subsection{Black-hole excision boundary conditions}
\label{sec:quasi-equil-bound}

In this paper, we are interested in the situation in which each black
hole is in quasiequilibrium and the boundary conditions required to
achieve this were worked out in Refs.~\cite{Cook-2002} and
\cite{cook-pfeiffer-2004a}.  The assumptions are essentially the same
as those required of an ``isolated horizon''(cf.\ 
\cite{ashtekar-etal-2000a,dreyer-etal-2003,ashtekar-krishnan-2003a,Jaramillo-etal:2004}).
To ensure that the black hole is in quasiequilibrium, we enforce the
following conditions.  First, we demand that the expansion
$\Oexpansion$, of the outgoing null rays, $\ONull^\mu$, vanish on the
excision surface, ${\cal S}$, thus forcing the boundary to be an
apparent horizon:
\begin{equation}\label{eq:expansion_on_S}
\Oexpansion\big|_{\cal S}=0.
\end{equation}
Next, we require that the shear $\Oshear_{\mu\nu}$ of the outgoing
null rays also vanish on the excision boundary,
\begin{equation}\label{eq:shear_on_S}
\Oshear_{\mu\nu}\big|_{\cal S}=0.
\end{equation}
In the absence of matter on ${\cal S}$,
Eqs.~(\ref{eq:expansion_on_S}) and (\ref{eq:shear_on_S}) are
sufficient to imply that
\begin{equation}\label{eq:dexpansion_on_S}
\LieD\ONull\Oexpansion\big|_{\cal S}=0.
\end{equation}
That is, initially, the apparent horizon will evolve along
$\ONull^\mu$.

Conditions (\ref{eq:expansion_on_S}),~(\ref{eq:shear_on_S})
and~(\ref{eq:dexpansion_on_S}) are coordinate independent, however the
final demand breaks precisely this coordinate freedom.  This final
condition is that the {\em coordinate location} of the apparent
horizon does not move initially in an evolution of the initial data.
Let the excision boundary surface, ${\cal S}$, be a spacelike
2-surface with topology $S^2$ and define $\Bnormal^i$ to be the
outward pointing unit vector normal to the surface.  If we let
\begin{equation}
\ONull^\mu \equiv \mbox{$\frac1{\sqrt{2}}$}(\Snormal^\mu + \Bnormal^\mu) 
\label{eq:BNullV_def}
\end{equation}
represent the set of outgoing null rays associated with ${\cal S}$,
and with the time vector written as
\begin{equation}
t^\mu = \Lapse\Snormal^\mu + \Shift^\mu,
\label{eq:time_vec_def}
\end{equation} 
then this condition can be expressed as
\begin{equation}\label{eq:t.k=0}
t^\mu \ONull_\mu\Big|_{\cal S}=0.
\end{equation}
This immediately yields
\begin{equation}\label{eq:Shift-perp-raw}
\Lapse\big|_{\cal S}=\Shift^i\Bnormal_i\big|_{\cal S}.
\end{equation}
To further analyze Eq.~(\ref{eq:Shift-perp-raw}), we split 
the shift vector into its component normal to the surface,
$\perpShift$, and a vector tangent to the surface, $\parShift^i$,
defined by
\begin{eqnarray}
\perpShift &\equiv& \Shift^i\Bnormal_i,
\label{eq:perpShift_def} \\
\parShift^i &\equiv& \BMetric^i_j\Shift^j,
\label{eq:parShift_def}
\end{eqnarray}
where $\BMetric_{ij}\equiv\SMetric_{ij} - \Bnormal_i\Bnormal_j$ is the metric induced on ${\cal S}$ by $\SMetric_{ij}$.
We see that Eq.~(\ref{eq:Shift-perp-raw}) is a
condition on the normal component of the shift, 
\begin{equation}
\perpShift\big|_{\cal S} = \Lapse\big|_{\cal S}.
\label{eq:perpShift_BC}
\end{equation}

The conformal transformation on $\SMetric_{ij}$, 
Eq.~(\ref{eq:Metric_decomp}), induces a natural conformal weighting for
$\BMetric_{ij}$ and for the unit normal to ${\cal S}$,
\begin{eqnarray}
\BMetric_{ij} &\equiv& \CF^4\CBMetric_{ij},
\label{eq:CBMetric_def} \\
\Bnormal^i &\equiv& \CF^{-2}\CBnormal^i.
\label{eq:CBnormal_def}
\end{eqnarray}
In terms of these conformal quantities, the condition in
Eq.~(\ref{eq:expansion_on_S}) takes the form of a nonlinear
Robin-type boundary condition on the conformal factor $\CF$,
\begin{align}\nonumber
\CBnormal^k\CCD_k\ln\CF\Big|_{\cal S} = 
-\frac14\bigg(&\CBMetric^{ij}\CCD_i\CBnormal_j + \frac{\TrExCurv}{6}\CF^2 \\
&               - \frac{\CF^{-4}}{8\CLapse}\CBnormal_i\CBnormal_j \big[\CLD{\Shift}^{ij}-\CMtd^{ij}\big]\bigg)
\bigg|_{\cal S}.
\label{eq:AH_BC}
\end{align}

We now turn our attention to Eq.~(\ref{eq:shear_on_S}), which can be
rewritten as
\begin{equation}\label{eq:Killing_shift_raw}
\CBCD^{(i}\parShift^{j)}\Big|_{\cal S}
- \mbox{$\frac12$}\CBMetric^{ij}\CBCD_k\parShift^k\Big|_{\cal S} = 
\mbox{$\frac{1}{2}$}\left(\CBMetric^i{}_k\CBMetric^j{}_l-\mbox{$\frac{1}{2}$}\CBMetric^{ij}\CBMetric_{kl}\right)\CMtd^{kl}\Big|_{\cal S},
\end{equation}
where $\CBCD^i$ denotes the covariant derivative compatible with
$\CBMetric_{ij}$.  Below, we will assume $\CMtd_{ij}\!=\!0$; in that case,
Eq.~(\ref{eq:Killing_shift_raw}) implies that $\parShift^i$ must be a
conformal Killing vector of the conformal metric, $\CBMetric_{ij}$ 
on the boundary ${\cal S}$:
\begin{equation}
\CBCD^{(i}\parShift^{j)}\Big|_{\cal S}
- \mbox{$\frac12$}\CBMetric^{ij}\CBCD_k\parShift^k\Big|_{\cal S} = 0.
\label{eq:Killing_shift}
\end{equation}
In practice, we write the parallel components of the shift as
\begin{equation}
\parShift^i = \Omega_r \xi^i,
\label{eq:parShift_BC}
\end{equation}
where $\xi^i$ is a rotational conformal Killing vector on
$\CBMetric_{ij}$ with affine length of $2\pi$, and $\Omega_r$ is a
constant.  As shown in Ref.~\cite{cook-pfeiffer-2004a}, the choice of
$\parShift^i$ directly parameterizes the spin of the associated
black hole.  The restriction on $\parShift^i$ is quite remarkable.
Regardless of the choice of the conformal metric $\CMetric_{ij}$ (and
thus for any $\CBMetric_{ij}$), Eq.~(\ref{eq:Killing_shift}) still
allows sufficient freedom to allow for the parameterization of a
rotation about any direction (by the choice of $\xi^i$) and with any
magnitude (by the choice of $\Omega_r$)\cite{cook-pfeiffer-2004a}.

To summarize, the quasiequilibrium conditions defined in
Eqs.~(\ref{eq:expansion_on_S}), (\ref{eq:shear_on_S}),
and~(\ref{eq:t.k=0}) define boundary conditions on the conformal
factor, $\CF$, via Eq.~(\ref{eq:AH_BC}) and on the shift vector,
$\Shift^i$, via Eqs.~(\ref{eq:perpShift_BC})
and~(\ref{eq:parShift_BC}).  These total to four of the five
necessary boundary conditions for solving the coupled elliptic
equations associated with the conformal thin-sandwich equations.
Missing is a condition on the conformal lapse, $\CLapse$.

However, as clearly shown in Ref.~\cite{cook-pfeiffer-2004a}, the
excision boundary condition on the lapse is intimately associated with
a degeneracy in the choice of the initial slicing condition.  In fact,
for stationary black-hole spacetimes, it is the choice of the lapse on
the excision boundary that uniquely fixes a particular initial slice.
For initial data representing systems that are nearly stationary, that
is systems in quasiequilibrium, it has also been shown that the choice
of the excision boundary condition for the lapse is largely
irrelevant.  Following Ref.~\cite{cook-pfeiffer-2004a}, we will choose
the excision boundary condition for the lapse from a set of convenient
and rather generic conditions.  Assuming that the excision boundary
is spherical, we use
\begin{subequations}\label{eq:Lapse-BCs}
\begin{align}
\label{eq:Lapse-BC-1}
\frac{d\Lapse\CF}{dr}\bigg|_{\cal S}&=0,\\
\label{eq:Lapse-BC-2}
\frac{d\Lapse\CF}{dr}\bigg|_{\cal S}&=\frac{\Lapse\CF}{2r}\bigg|_{\cal S},\\
\label{eq:Lapse-BC-3}
\Lapse\CF\big|_{\cal S}&=\frac{1}{2}.
\end{align}
\end{subequations}

\subsection{Quasi-circular orbits}
\label{sec:quasi-circ-orbits}

The formalism reviewed so far in Secs.~\ref{sec:conf-thin-sandw}
and~\ref{sec:quasi-equil-bound} provides a very general framework for
constructing initial data for black-hole binaries.  Let us now make a
specific choice for part of the freely-specifiable initial data.
Consider the choice $\CMtd_{ij} \equiv \dtime\CMetric_{ij}=0$.  This
choice implies that the conformal three-geometry is, at least
momentarily, stationary with respect to the time vector, ${\mathbf
t}$, given by Eq.~(\ref{eq:time_vec_def}).  Our choice for the
boundary condition on the shift at infinity given in
Eq.~(\ref{eq:shift_BC_infty}) implies that our time vector has a {\em
helical} form where the amount of ``twist'' in the vector is
parameterized by $\Omega_0\equiv|{\mathbf\Omega_0}|$.

If we assume that ${\mathbf t}$ is an approximate Killing vector of
the spacetime, then $\CMtd_{ij}=0$ is a direct consequence.  With a
proper choice of $\Omega_0$, the resulting initial data represents a
binary system where the black holes are in quasi-circular orbits.
This statement begs two questions: 1) How do we determine the proper
choice of $\Omega_0$ so as to obtain quasi-circular orbits?  2) How do
we interpret the initial data for other choices of $\Omega_0$?

An answer to the first question was proposed by Gourgoulhon et~al.\cite{gourgoulhon-etal-2002a} where they made the ansatz that
$\Omega_0$ be chosen so that the ADM mass\cite{ADM} and the Komar
mass\cite{komar59} agree.  We call this the ``Komar-mass ansatz'', and
when it is applied, we refer to it as the ``Komar-mass condition''.  

The ADM energy is an invariant definition of the total energy of a
spacetime as measured by an inertial observer at spacelike infinity.
The ADM energy is written as a surface integral at infinity
\begin{equation}\label{eq:EADM}
E_{\mbox{\tiny ADM}}=\frac{1}{16\pi}\oint_{\infty} 
  \SCD_j\left({\cal G}^j_i - \delta^j_i {\cal G}\right)d^2S^i,
\end{equation}
where ${\cal G}_{ij}\equiv \SMetric_{ij} - f_{ij}$, ${\cal
G}=\SMetric^{ij}{\cal G}_{ij}$, and $f_{ij}$ is a flat metric.  If the
total linear momentum of the system (as measured by the same inertial
observer at infinity) vanishes, then the ADM energy is usually referred
to as the ADM mass.  In our notation, the Komar mass can be written as
\begin{equation}\label{eq:M_Komar}
M_{\mbox{\tiny K}}=\frac{1}{4\pi}\oint_{\infty}
  \left(\SCD_i\Lapse-\Shift^j\ExCurv_{ij}\right)d^2S^i.
\end{equation}
The Komar mass is a valid expression for the total mass of a system
only when the system possesses a global timelike Killing vector so that
the system is stationary.  It is therefore quite reasonable to assume
that by choosing $\Omega_0$ so that the Komar-mass ansatz will yield
initial data which is nearly stationary (i.e.\ in quasiequilibrium),
which would require that the black holes be in quasi-circular orbits.

The effectiveness of the Komar-mass ansatz has been tested numerically
in black-hole initial data by comparison with post-Newtonian data for
binaries in circular orbits\cite{Damour-etal-2002,cook-pfeiffer-2004a}
(and for neutron-star binaries\cite{taniguchi-gourgoulhon-2002}).
There are also additional theoretical reasons for expecting that
configurations satisfying the Komar-mass condition ($E_{\mbox{\tiny
ADM}}=M_{\mbox{\tiny K}}$) will represent systems in quasiequilibrium
(cf.\ Refs.~\cite{friedman-etal-2002,gourgoulhon-etal-2002a} and
references therein).  These arguments show that a system in
quasiequilibrium necessarily satisfies the Komar mass ansatz.  However, as far
as we know, there is no guarantee that a system satisfying the
Komar-mass condition is necessarily in quasiequilibrium.  It would
therefore be interesting to compare the Komar-mass ansatz against an
independent method for determining circular orbits.  The
effective-potential method\cite{cook94e} will be used below to provide
such a comparison.

It is worth noting that when using the Komar-mass ansatz in the
context of binary systems, care must be used in evaluating
Eq.~(\ref{eq:M_Komar}).  An observer moving along the approximate
helical Killing vector ${\mathbf t}$ is not an inertial observer.  For
a true stationary spacetime, the helical Killing vector can be split
globally into separate timelike and rotational Killing vectors and the
timelike Killing vector is used to define the timeline of the inertial
observer making the measurements.  The approximate helical Killing
vector cannot be split globally, however it can be split {\em
asymptotically} at spacelike infinity.  In evaluating the
Eq.~(\ref{eq:M_Komar}), it is necessary to remove the
${\mathbf\Omega_0\times{\bf r}}$ term from the shift, $\Shift^i$ so
that the same inertial observer is used to evaluate both
$M_{\mbox{\tiny K}}$ and $E_{\mbox{\tiny ADM}}$.  Often, the Komar
mass is written in a from similar to that of Eq.~(\ref{eq:M_Komar}),
but with the term involving the shift absent.  In many cases, this
yield a correct expression for the Komar mass since the contraction of
the shift with the extrinsic curvature falls off faster than $1/r^2$
when the ${\mathbf\Omega_0\times{\bf r}}$ contribution to the shift is
omitted.  However, it is {\em not} always correct to simply drop this
term.  For example, in Painlev\'e-Gullstrand
coordinates\cite{painleve-1921,gullstrand-1922,martel-poisson-2001}
and their extension to the full Kerr-Newman spacetime\cite{doran-2000},
this term contains {\em the entire contribution to the Komar mass}.

In order to answer the second question of how to interpret the initial
data when $\Omega_0$ is no longer chosen via the Komar-mass condition, we
must no longer think of the helical time vector as an approximate
Killing vector of the spacetime.  A more general interpretation of the
choice $\CMtd_{ij} \equiv \dtime\CMetric_{ij}=0$ is that the
``velocity'' of the conformal three-geometry vanishes on the
initial-data slice.  In the context of a binary configuration this
suggests that the system is at either pericenter or apocenter of some
general bound or unbound orbit.  When $\Omega_0=0$, the system will
have no orbital angular momentum and will represent a generalized
version of Misner\cite{misner63} or
Lindquist\cite{lindquist63,brill-lindquist-1963} initial data.

\subsection{Corotating and non-spinning black-hole binaries}
\label{sec:corot-non-rotat}

In constructing black-hole binary initial data, we will certainly need
the ability to specify the spins of the individual black holes.  In
Sec.~\ref{sec:quasi-equil-bound}, we mentioned that the excision
boundary condition on the parallel components of the shift,
$\parShift^i$, can be used to set the spin of each black hole.  It is
tempting to want to interpret $\Omega_r$ in Eq.~(\ref{eq:parShift_BC})
as the rotational angular velocity of the black hole.  However this is
not the case.  To understand the role of $\parShift^i$ in determining
the spin on the black holes, it is useful to consider the special
cases of corotating and non-spinning black holes.  While neither
case is expected to be seen astrophysically, these cases are
useful because they represent situations where we know either what the
boundary condition should be or what the final spin should be.

Corotating black hole binaries represent the case where the black
holes are rotating synchronously with the orbital motion.  A great
deal of attention has been payed to such binaries because they
represent the only configuration of two black holes that can possess a
{\em true} helical Killing
field\cite{detweiler92,detweiler94,Whelan-etal-2000,Whelan-etal-2002,friedman-etal-2002,Klein-2005}.
A serious fault with such spacetimes is that they cannot be
asymptotically flat since they contain a balancing amount of incoming
and outgoing radiation for all time.  However, for our purposes, they
are ideal.  First, because we at most have only an approximate helical
Killing vector, we can still have an asymptotically flat solution.
More importantly, we know that the corotating case possess a Killing
horizon.  This means that the proper choice for the parallel
components of the shift is unambiguously given by $\parShift^i=0$.

Since corotating black holes necessarily have some non-vanishing
rotation (and therefore spin) as seen by an asymptotic inertial
observer, clearly $\Omega_r$ cannot represent the rotational 
angular velocity of the black hole.  At least for a Newtonian
binary system in corotation, the rotational angular velocity
should equal the orbital angular velocity of the binary system.
For post-Newtonian computations of corotating black-hole binaries,
this is, in fact, the condition used to set the spins of the 
black holes\cite{Damour-etal-2002,mora-will-2004}.  To leading
order, this is correct.  However, as we will demonstrate below,
higher order corrections are needed to correctly estimate the
rotational angular velocity of corotating black holes.

Non-spinning black-hole binaries are, in some sense, the simplest
case.  Here, the individual black holes have no spin as measured by an
asymptotic inertial observer.  In previous works, we have referred to
such systems as having ``irrotational'' black
holes\cite{Cook-2002,cook-pfeiffer-2004a}.  Perhaps, however, this is
not the best terminology as it carries with it the notion of fluid
motion which is not appropriate for black holes.  From now on, we will
refer to such black holes as being ``non-spinning''.  How do we choose
$\parShift^i$ to yield black holes with no spin?  We argued
previously\cite{cook-pfeiffer-2004a} that for a binary with orbital
angular velocity $\Omega_0$, we should choose $\Omega_r=\Omega_0$.
Also, for spherical excision boundaries, we chose $\xi_i$ as the flat
space rotational Killing vector, projected into the excision surface,
that generates a rotation about an axis parallel to the orbital
angular momentum vector.  We also showed that this condition lead to
reasonable results.  To leading order this is correct, but again we
will show below that it is necessary to modify this choice for the
boundary condition in order to produce non-rotating black-hole
binaries.

\subsection{Conformally flat maximally sliced models}
\label{sec:conf-flat-maxim}

So far, we have discussed the approach used to construct binary
initial data in rather general terms.  We have discussed the conformal
thin-sandwich decomposition and the choice of boundary conditions for
constrained variables within that formalism.  These constrained
variables ($\CF$, $\Shift^i$, $\CLapse$) can only be determined after
values for the unconstrained variables ($\CMetric_{ij}$, $\CMtd_{ij}$,
$K$, $\dtime{K}$) have been chosen.  In
Sec.~\ref{sec:quasi-circ-orbits}, we discussed the interpretation of
initial data when the choice $\dtime\CMetric_{ij}\equiv\CMtd_{ij}=0$
was made.  We will use this choice for $\CMtd_{ij}$ in all models
constructed below.

We will also make the assumption of {\em conformal flatness} in all of
the models we construct.  That is, we will choose $\CMetric_{ij}$ to
be a flat metric.  This choice impacts upon the physical content of
the initial data that we will construct.  It is well known that the
spatial metric for a relativistic binary system cannot be conformally
flat\cite{rieth-Math-Grav}, nor can the spatial metric of a stationary
spinning black hole\cite{garat-price-2000,Kroon:2004} or even the
metric of a boosted black hole.  However, the errors introduced by the
assumption of conformal flatness are not ``grave''.  There are efforts
underway to improve the choice of the conformal metric.  Some of these
efforts involve using an analytic metric obtain from post-Newtonian
theory (cf. Ref.~\cite{tichy-etal-2003}).  However a more
self-consistent approach has been developed by Shibata et~al.\cite{Shibata-etal-2004,Uryu-etal-2006} for the case of
neutron-star binaries and it should be possible to adapt this approach
to black-hole binaries in the future.

Finally, we choose to use maximal slicing,
$\TrExCurv=\dtime\TrExCurv=0$, in all of the models we construct.
Which member of the family of possible maximal slices we choose will
depend upon our choice for the boundary condition for the
lapse\cite{cook-pfeiffer-2004a}.  As a slicing condition, we do not
expect this choice to impact significantly upon the physical content
of the initial data.

\subsection{Numerical code}
\label{sec:numerical-code}

We solve the elliptic equations of the conformal thin-sandwich
decomposition using the pseudo-spectral collocation method described
in Refs.~\cite{Pfeiffer2003,cook-pfeiffer-2004a}.  As usual with spectral
methods (see, e.g. \cite{Boyd:2001}), the solution is expressed as a
truncated series of basis-functions, and is represented by a set of
expansion coefficients.  For appropriate basis-functions, discretization
errors decay exponentially with the number of retained
basis-functions~\cite{Boyd:2001}.  The elliptic equations take the
form of a set of nonlinear algebraic equations for the expansion
coefficients.  These algebraic equations are solved with a
Newton-Raphson method, where in each step a linear problem is solved
via standard Krylov subspace techniques~\cite{Templates} like
preconditioned fGMRES~\cite{Saad:1993}.

For the binary black hole solutions of this paper, we need to solve
elliptic equations in a computational domain with two excised spheres.
To do so, the computational domain is split into smaller subdomains,
namely spherical shells and rectangular blocks. A shell is placed
around each excision surface and a third shell, using a compactified
radial coordinate, extends from some intermediate radius surrounding
both holes out to the outer boundary which is typically placed at a
radius of $\sim 10^8$.  The space between is filled by a collection of
rectangular blocks.

Further details of the numerical code can be found in
Ref.~\cite{cook-pfeiffer-2004a}.  In particular, Fig.~6 of
Ref.~\cite{cook-pfeiffer-2004a} displays the convergence of the
code for a typical configuration.  The calculations below were
performed at a resolution comparable to $N=60$ in this figure and
correspond to discretization errors on the order of $10^{-5}$ or
$10^{-6}$ for most of the quantities we consider.

\section{Computing the quasi-local spin of a black hole}
\label{sec:comp-quasi-local}

In the previous discussion, we referred to the spin of an individual
black hole in a binary system.  However, there is no unique, rigorous
definition of the spin of a black hole unless it is in isolation and
stationary.  We can, however, rigorously define the total angular
momentum of an initial-data slice as measured by an inertial observer
at infinity.  Usually referred to as the ADM angular momentum, it
can be expressed as
\begin{equation}
\label{eq:JADM}
J_{(\xi)}=\frac{1}{8\pi}\oint_{\infty}
  \left(\ExCurv_{ij}-\SMetric_{ij}\TrExCurv\right)\xi^j\,d^2S^i,
\end{equation}
where $\xi^i$ is a Killing vector of $\SMetric_{ij}$.  For
asymptotically flat data, we can choose this to be each of the three
flat-space rotational Killing vectors in order to determine the three
components of the total angular momentum of the system.

Although there is no unique, rigorous definition for the spin of an
individual black hole in general, we can estimate the spin based upon
a quasi-local definition.  There are many different quasi-local
definitions for spin\cite{Szabados-2004}.  While motivated in
different ways, they tend to take on a similar form within a $3+1$
framework.  One of the earliest useful definitions was derived by
Brown and York\cite{brown-york-1993}.  This same definition was
rederived more recently within the {\em isolated} and {\em dynamical
horizons} framework of Ashtekar and
Krishnan\cite{ashtekar-krishnan-2003a,Ashtekar-Krishnan-2004}.  In
either case, the quasi-local spin can be expressed as
\begin{equation}
\label{eq:QL_spin0}
S_{(\xi)}=\frac{1}{8\pi}\oint_{\cal S}
  \left(\ExCurv_{ij}-\SMetric_{ij}\TrExCurv\right)\xi^j\,d^2S^i,
\end{equation}
where now $\xi^i$ is a Killing vector of $\BMetric_{ij}$.  We
note the remarkable similarity between Eqs.~(\ref{eq:JADM}) and
(\ref{eq:QL_spin0}).  The main difference between them is that the
former is evaluated at infinity while the latter is evaluated on the
apparent horizon.  For the initial data we are constructing, this will
be the excision surface.  In terms of the variables used in the
conformal thin-sandwich decomposition, we evaluate the spin as
\begin{equation}
\label{eq:QL_spin}
S_{(\xi)}=\frac{1}{16\pi}\oint_{\cal S}
  \frac{1}{\CLapse}\left[\CLD{\Shift}_{ij} - \CMtd_{ij}\right]
  \xi^j\CBnormal^i\sqrt{\CBMetric}\,d^2x,
\end{equation}
where we have used the fact that $\xi^i$ is tangent to the excision
surface, i.e. $\CMetric_{ij}\xi^i\CBnormal^j=0$.

The only choice that must be made in evaluating the spin is to choose
the Killing vector $\xi^i$.  The problem is that, in general, an exact
Killing vector will not exist.  However, there are two reasonable
choices that can be made.  If our excision surface (apparent
horizon) is a coordinate sphere within our flat conformal geometry,
then we can choose to approximate $\xi^i$ by one of the 3 flat-space
rotational Killing vectors centered on the excision sphere and
projected onto its surface.  We will denote these three choices
by
\begin{subequations}\label{eq:CKVs}
\begin{align}
\label{eq:CKV_x}
\xi^i_{(x)} &= x^j_s\epsilon^{ixj}, \\
\label{eq:CKV_y}
\xi^i_{(y)} &= x^j_s\epsilon^{iyj}, \\
\label{eq:CKV_z}
\xi^i_{(z)} &= x^j_s\epsilon^{izj},
\end{align}
\end{subequations}
where we assume Cartesian coordinates and ${\mathbf x}_s$ is measured
relative to the center of the excision sphere.

An alternative is to attempt to solve Killing's equation.  Once we have
solved the constraint equations and have full initial data, we know
$\BMetric_{ij}$, the physical metric projected onto the excision
surface ${\cal S}$.  In Ref.\cite{dreyer-etal-2003}, Dreyer et~al.\ outlined a general method for finding the Killing vectors on a
closed 2-surface.  We give the details of our implementation of this
method in Appendix~\ref{sec:solv-kill-equat}.  Here, we simply note
that exact solutions of Killing's equation will not exist in general
so what we find are ``approximate'' Killing vectors.

It is difficult to make a meaningful quantitative measure of how
far a given solution deviates from being a true Killing vector.
For our purposes, we will attempt to gauge the accuracy of our
measured spins by comparison with expected values.  For the case
of corotating black holes, we know that our boundary condition on
$\parShift^i$ is correct and we know, at least to leading order,
what the spin of a corotating black hole should be in a binary.
Thus, we can compare our quasi-local definition for the spin of
a black hole against an analytic result.

Below, when we discuss the computed quasi-local spin, we will use
the following notation for simplicity:
\begin{equation}\label{eq:Spin-defs}
\begin{array}{rcl}
  S_x &:& \mbox{computed using Eq.~(\ref{eq:CKV_x})} \\
  S_y &:& \mbox{computed using Eq.~(\ref{eq:CKV_y})} \\
  S_z &:& \mbox{computed using Eq.~(\ref{eq:CKV_z})} \\
  S_K &:& \left\{\begin{array}{l}\mbox{computed using an approximate} \\ 
                           \mbox{solution of Killing's equation.}
          \end{array}\right.
\end{array}
\end{equation}

\section{Corotating Binaries}\label{sec:corotating-binaries}

Although corotation is not considered to be an astrophysically
realistic state for black-hole binaries, it is an important test case
because it is the one configuration for black-hole binaries that is
compatible with a true helical Killing
vector\cite{detweiler92,detweiler94,friedman-etal-2002}.  The
thermodynamic relations obtained by Friedman et~al.\cite{friedman-etal-2002},
\begin{equation}\label{eq:thermo_ident0}
  \delta E_{\mbox{\tiny ADM}} = \Omega_0\delta J_{\mbox{\tiny ADM}} 
  + \sum{\kappa_i\delta {\cal A}_i},
\end{equation}
should apply to our conformally-flat data if we had a true
helical Killing vector and we assume no local change in entropy.  In
Eq.~(\ref{eq:thermo_ident0}), $J_{\mbox{\tiny ADM}}$ 
is the magnitude of the total ADM angular momentum of the system,
$\kappa_i$ and ${\cal A}_i$ are respectively the surface gravity and
area of the Killing horizon of each black hole.  Corotating black
holes also have a non-vanishing spin and, more importantly, a
physically well-defined notion of the rate of rotation.  Together,
these give us a firm analytic foundation against which we can test our
initial data.

Let us first consider the thermodynamic identity in
Eq.~(\ref{eq:thermo_ident0}).  We are free to define a fundamental
length scale for each of our initial-data solutions.  We can use this
freedom to scale our solutions in an attempt to have it satisfy this
identity.  Of course, Eq.~(\ref{eq:thermo_ident0}) allows for too much
variation for a single length rescaling to guarantee the enforcement
of the identity in general.  However, we can use the freedom to define
a fundamental length scale $\chi(s)$ along a sequence of initial data
to enforce
\begin{equation}
  \delta{E_{\mbox{\tiny ADM}}} = \Omega_0\, \delta{J_{\mbox{\tiny ADM}}}
\label{eq:thermo_identity}.
\end{equation}
This approach has been discussed
previously\cite{gourgoulhon-etal-2002b,cook-pfeiffer-2004a}, but 
for completeness, we cover it again here.

Using our freedom to define the fundamental length scale $\chi(s)$
along a sequence we define the dimensionful total energy
$E_{\mbox{\tiny ADM}}(s)$, total angular momentum $J_{\mbox{\tiny
ADM}}(s)$, and orbital angular velocity $\Omega_0(s)$ consistently
via,
\begin{align}
E_{\mbox{\tiny ADM}}(s) &\equiv \chi(s)e(s),
\label{eq:energy_scale} \\
J_{\mbox{\tiny ADM}}(s) &\equiv \chi^2\!(s)j(s),
\label{eq:angmom_scale} \\
\Omega_0(s) &\equiv \chi^{-1}\!(s)\omega(s).
\label{eq:angvel_scale}
\end{align}
Then, to enforce Eq.~(\ref{eq:thermo_identity}), it is sufficient to
determine the change in $\chi(s)$ between two points on the sequence.
This can be done by integrating along the sequence from a point $s_1$
to another point $s_2$.  Doing so, we find\footnote{Note that the
similar equation in Ref.\cite{cook-pfeiffer-2004a} contains a factor
of 2 error in the denominator of the integrand.}
\begin{equation}
  \chi(s_2) = \chi(s_1)\exp\left\{-\int_{s_1}^{s_2}{
    \frac{e^\prime(s) - \omega(s)j^\prime(s)}{e(s) -2\omega(s)j(s)}}ds\right\},
\label{eq:thermo_scale}
\end{equation}
where a prime denotes differentiation along the sequence.

Now, if the Komar-mass ansatz is reasonable, then a sequence of
initial data with varying separation that satisfies this ansatz should
represent a binary in a nearly adiabatic evolutionary sequence of
quasi-circular orbits that satisfy Eq.~(\ref{eq:thermo_ident0}).  But,
having scaled the data to satisfy Eq.~(\ref{eq:thermo_identity}), the
identity in Eq.~(\ref{eq:thermo_ident0}) has been reduced to
$\sum\kappa_i\delta{\cal A}_i=0$.  To leading order, $\kappa=1/4M_{\rm irr}$
for each hole.  Together with the definition of the irreducible mass
\begin{equation}\label{eq:Mirr_def}
M_{\rm irr}=\sqrt{\frac{A_{AH}}{16\pi}},
\end{equation}
where $A_{AH}$ is the area of the black hole's apparent horizon, we
find $\kappa\delta{\cal A} \approx 8\pi\delta{M_{\rm irr}}$ for each
black hole.  So, if the Komar-mass ansatz is reasonable we should find
that $\sum(\delta{M_{\rm irr}})_i\approx0$ along a sequence of
initial-data sets.

The second column of Table~\ref{tab:thermo_irr_mass} lists the
irreducible mass of one corotating black hole as we vary the orbital separation
and enforce Eq.~(\ref{eq:thermo_identity}).  Given our excision
boundary condition, Eq.~(\ref{eq:perpShift_BC}), the apparent horizon
is, at least instantaneously, a Killing horizon.  The only deviation
we should expect in $M_{\rm irr}$ should be due to the fact that we do
not have a true helical Killing vector and that the apparent horizon
is not part of a global Killing horizon.  As we see from the table,
$M_{\rm irr}$ changes only by about one part in $10^4$ over the
entire sequence of separations.  While small, the variations seen in
$M_{\rm irr}$ are larger than the level of accuracy of the initial
data computations and the leading digits of the variation are also
above the level of truncation error that result from integrating
Eq.~(\ref{eq:thermo_scale}).  Thus, the variation in $M_{\rm irr}$
seems to be a true artifact of our approximate helical symmetry.  On
the other hand, it is quite small and this lends support to the
Komar-mass ansatz.

\begin{table}
\caption{The irreducible mass $M_{\rm irr}$ of a single black hole in
a corotating, non-spinning, and leading-order (LO) non-spinning
binary as we vary the coordinate separation $d$.  The length scale is
set so that the ADM mass of the binary at very large separation is 1.
Lapse boundary condition (\ref{eq:Lapse-BC-2}) was used for all data.}
\begin{ruledtabular}
\begin{tabular}{l|cc|c}
 & \multicolumn{3}{c}{$M_{\rm irr}$} \\
$d$ & Corotation & Non-spinning & ``LO'' non-spinning \\
\hline
40   & 0.5000000 & 0.5000000 & 0.5000000 \\
35   & 0.5000001 & 0.5000001 & 0.4999988 \\
30   & 0.5000002 & 0.5000002 & 0.4999966 \\
25   & 0.5000005 & 0.5000005 & 0.4999902 \\
20   & 0.5000010 & 0.5000011 & 0.4999727 \\
19   & 0.5000013 & 0.5000014 & 0.4999659 \\
18   & 0.5000016 & 0.5000017 & 0.4999568 \\
17   & 0.5000020 & 0.5000021 & 0.4999446 \\
16   & 0.5000025 & 0.5000026 & 0.4999281 \\
15   & 0.5000032 & 0.5000033 & 0.4999050 \\
14.5 & 0.5000037 & 0.5000037 & 0.4998901 \\
14   & 0.5000042 & 0.5000042 & 0.4998721 \\
13.5 & 0.5000048 & 0.5000048 & 0.4998504 \\
13   & 0.5000056 & 0.5000055 & 0.4998238 \\
12.5 & 0.5000065 & 0.5000064 & 0.4997911 \\
12   & 0.5000076 & 0.5000074 & 0.4997504 \\
11.5 & 0.5000090 & 0.5000086 & 0.4996992 \\
11   & 0.5000107 & 0.5000102 & 0.4996340 \\
10.5 & 0.5000129 & 0.5000120 & 0.4995498 \\
10   & 0.5000157 & 0.5000144 & 0.4994393 \\
9.5  & 0.5000193 & 0.5000173 & 0.4992918 \\
9    & 0.5000240 & 0.5000211 & 0.4990908 \\
8.5  & 0.5000308 & 0.5000259 & 0.4988091 \\
8    & 0.5000401 & 0.5000330 & 0.4984008 \\
7.5  & 0.5000529 & 0.5000423 & 0.4977815 \\
7    & 0.5000702 & 0.5000546 & 0.4967842 
\end{tabular}
\end{ruledtabular}
\label{tab:thermo_irr_mass}
\end{table}

\subsection{Effective-potential method}
\label{sec:effect-potent-meth}

Another way of testing the Komar-mass ansatz is to use an independent
method for identifying circular orbits.  An ``effective-potential''
method for determining circular orbits from sequences of initial-data
sets was outlined in Ref.~\cite{cook94e}.  This method is motivated by
variational techniques, but does not have a rigorous theoretical
foundation.  Never-the-less, it has been used successfully for
black-hole binary initial
data\cite{cook94e,baumgarte-2000,pfeiffer-etal-2000} and has been
shown to agree with the Komar-mass condition in the case of circular
orbits for thin shells of collisionless
matter\cite{skoge-baumgarte-2002}.  Agreement between the two methods
was also demonstrated, within error bars, at the location of the
inner-most stable circular orbit of sequences of circular orbits
produced using puncture data\cite{tichy-bruegmann-2004} (which uses
different and more simplifying assumptions).

The method is straightforward, but has one significant point of
ambiguity.  The effective-potential (EP) method identifies circular
orbits as configurations with a minima in the binding energy along
sequences of configurations where the total angular momentum is held
fixed.  The ambiguity is associated with the same freedom to define
the fundamental length scale mentioned above.  In Ref.~\cite{cook94e},
this ambiguity was resolved by demanding that the mass of each black
hole as defined by the Christodoulou formula\cite{christ70} be held
constant along sequences of configurations.

The Christodoulou formula
\begin{equation}
M^2 = M^2_{\rm irr} + \frac{S^2}{4 M^2_{\rm irr}},
\label{eq:christ_Mass}
\end{equation}
includes the spin of the black hole which adds another level of
complication.  The EP method was originally applied
only to configurations in which the direction and magnitude of the
spin was also held constant along sequences of configurations.  With
this restriction, holding the Christodoulou mass $M$ constant was
equivalent to holding the irreducible mass $M_{\rm irr}$ constant.
Furthermore, with this restriction on the spin, finding minima of the
binding energy was equivalent to finding minima of the ADM energy.

However, when we consider sequences of corotating binaries, the
magnitude of the spins of the black holes are no longer held constant
and we must reconsider how we will fix the mass freedom.  From
Friedman et~al.\cite{friedman-etal-2002} and the results shown
above, is seems clear that we should hold the irreducible mass
(or the area of the apparent horizon) fixed along sequences of
configurations.  We will define the total mass $m$ and reduced 
mass $\mu$ as
\begin{align}
  \label{eq:m_total}
  m &\equiv {M_{\rm irr}}_1 + {M_{\rm irr}}_2, \\
  \label{eq:m_reduced}
  \mu &\equiv \frac{{M_{\rm irr}}_1{M_{\rm irr}}_2}{m},
\end{align}
and then the binding energy $E_b$ by
\begin{equation}
  E_b \equiv E_{\mbox{\tiny ADM}} - m.
\label{eq:E_binding}
\end{equation}
With this definition of the binding energy, minima of the binding
energy and the ADM energy agree for corotating configurations as
well as for sequences with the spins held fixed (so long as we also
hold the individual irreducible masses fixed).

We will, therefore, adopt the following functional definition for the
EP method.  We will take as configurations with
circular orbits, those initial data configurations that have a minimum
of the ADM (or binding) energy along sequences of configurations where
the total angular momentum, irreducible masses of the black holes, and
direction of the black holes spins are held fixed.  This definition is
sufficient to handle both corotating and non-rotating black-hole
binaries.  The case of black-holes with arbitrary spin will be
considered in future investigations.

\begin{figure}
\includegraphics[width=\linewidth]{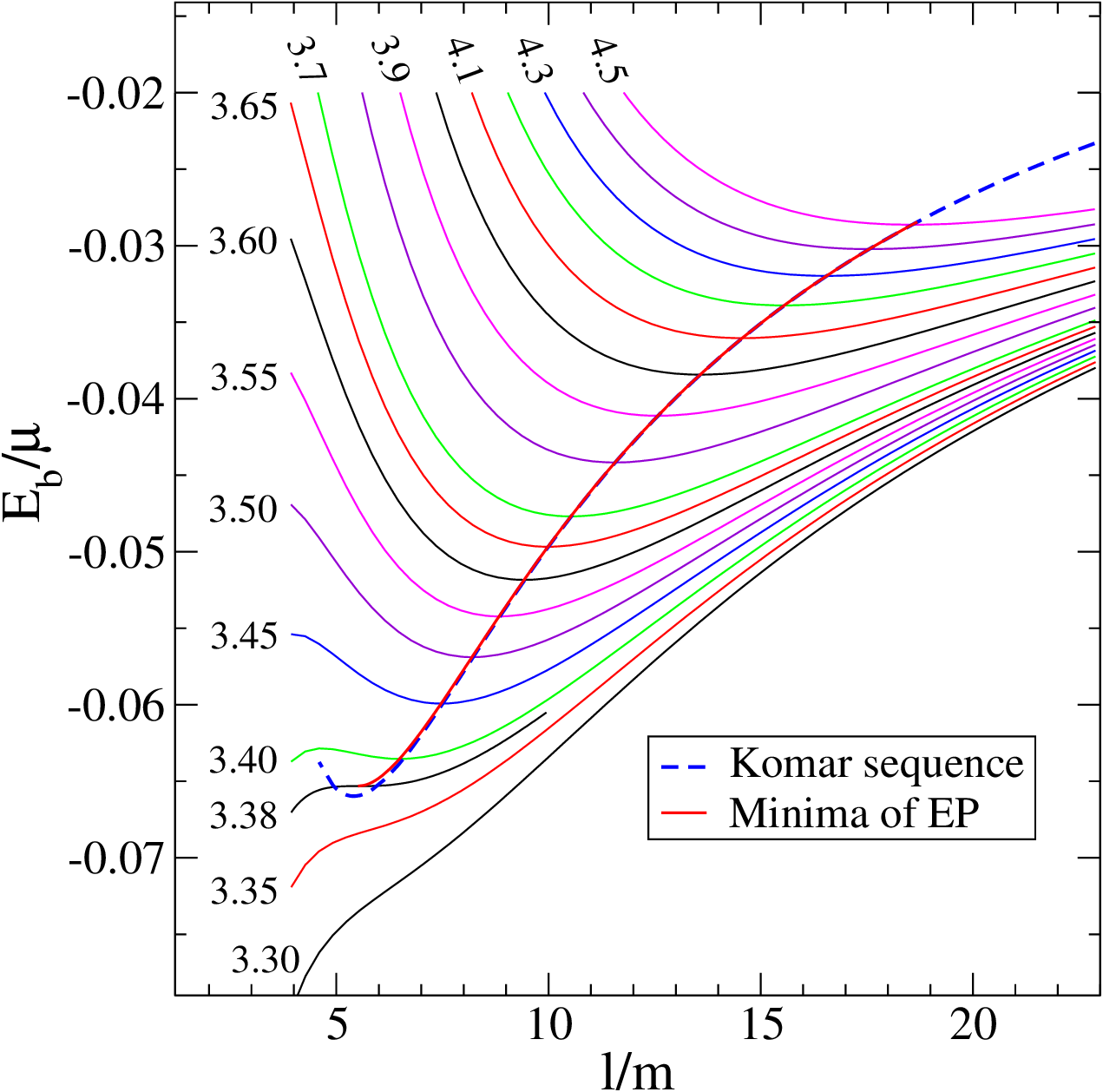}
\caption{\label{Fig:EP_corot}Effective potential (EP) curves $E_b/\mu$
for corotating black holes vs. separation $\ell/m$.  These curves are
labeled by the orbital angular momentum $J/\mu{m}$ which is kept
constant along each curve.  The thick red line connecting the minima of
the EP-curves represents circular orbits; it terminates at the
innermost stable circular orbit at the inflection point in the
EP-curve at $J/\mu{m}=3.38$.  Also plotted as a dashed blue line is the
sequence of circular orbits determined by the Komar-mass ansatz. }
\end{figure}

Figure~\ref{Fig:EP_corot} is a plot of the dimensionless binding
energy $E_b/\mu$ as a function of the dimensionless proper separation
$\ell/m$ between the apparent horizons for equal-mass corotating black
holes.  Each line of constant total angular momentum $J/\mu m$ is an
``EP curve'' and the local minimum of that curve represents the
circular-orbit configuration having that value of the total angular
momentum.  Passing through the set of minima, and plotted with a solid
line, is the EP sequence of circular orbits.  At small
separations, this sequence terminates at the inflection point on the
first EP curve that does not contain a local minimum.  This is the
short EP curve having
$J/\mu{m}\approx3.38$.  The inflection point in the EP curves marks the
termination of stable circular orbits and the configuration at the
point is referred to as the ``inner-most stable circular orbit''
(ISCO) configuration.  Also plotted in this figure with a dashed line
is the sequence of circular orbits as defined by the Komar-mass
ansatz.  For a Komar sequence, the ISCO configuration is defined as
the model with the minimum binding energy.  Notice that the Komar and
EP sequences are nearly coincident except for the regime near the
ISCO.

\begin{figure}
\includegraphics[width=\linewidth]{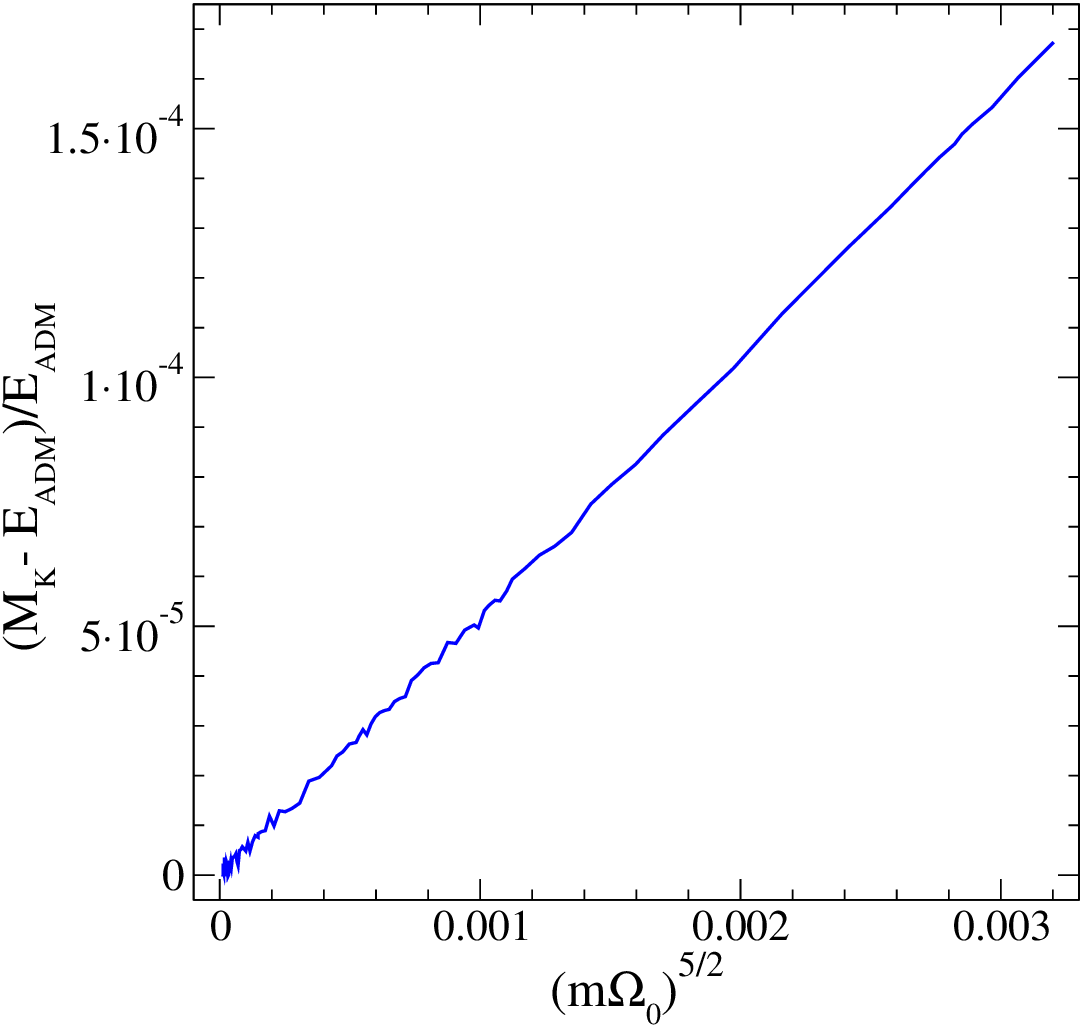}
\caption{\label{Fig:KomarErr_corot} Violation of the Komar-mass
condition when the effective potential method is used to determine the
sequence of circular orbits.  Here, corotating equal-mass binaries
are considered.  $m\Omega_0$ denotes the orbital angular frequency, so
that large separations correspond to small values of $m\Omega_0$.}
\end{figure}

A more quantitative comparison of the sequences obtained via the
Komar-mass ansatz and the EP method is found by examining the error in
the Komar-mass condition for circular orbits along a sequence of such
orbits defined by the EP method.  Figure~\ref{Fig:KomarErr_corot}
plots this error, $M_{\mbox{\tiny K}} - E_{\mbox{\tiny ADM}}$, scaled
relative to the ADM energy versus the $5/2$-power of the dimensionless
orbital angular velocity $m\Omega_0$.  Interestingly, we see that the
relative error is nearly linear in $(m\Omega_0)^{5/2}$.  We see that
the deviation is quite small, even at the ISCO near
$(m\Omega_0)^{5/2}\sim0.0032$, where the error is roughly $0.015\%$.
From the jaggedness of the curve at large separations 
(small $m\Omega_0$), it is evident
that the errors are nearing the level of truncation error for the
measurement of the energies.  However, the measured deviation is
clearly physical, not numerical.

We see that the Komar-mass condition and the EP method appear to agree
to a very high degree in determining configurations with circular
orbits.  But, there is a measurable difference.  As a final
comparison, we can rescale the sequence of circular orbits defined by
the EP method so as to satisfy Eq.~(\ref{eq:thermo_identity}) just as
was done for the sequence defined by the Komar-mass condition.  When we
examine the variation of $M_{\rm irr}$ along the rescaled EP sequence,
we find the largest deviations are more than a factor of 100 smaller
than those seen in the sequences defined by the Komar-mass condition
displayed in Table~\ref{tab:thermo_irr_mass}.  This level of
variability is consistent with the level of truncation error in the
numerics.  Thus, to the numerical precision of the calculations, we
find that the sequence of corotating equal-mass binaries in circular
orbits defined by the EP method satisfies the thermodynamic identity
of Eq.~(\ref{eq:thermo_ident0}).

\subsection{Spin}
\label{sec:CO-spin}

Another reason for considering corotating binaries is that we want to
calibrate our techniques for computing the quasi-local spin of a black
hole, and corotating binaries have a physically well-defined notion of
the rate of rotation of each black hole.  From a Newtonian
perspective, a corotating binary has the rotational angular velocities
of the individual objects equal to the orbital angular velocity.  But,
this is a non-local statement and, as such, is not well defined in the
context of general relativity.  A relativistically reasonable notion
of corotation might be to connect the rate of rotation of one black
hole with the local rate of rotation of tidal perturbations due to the
orbit of its companion.

\begin{figure}
\includegraphics[width=\linewidth]{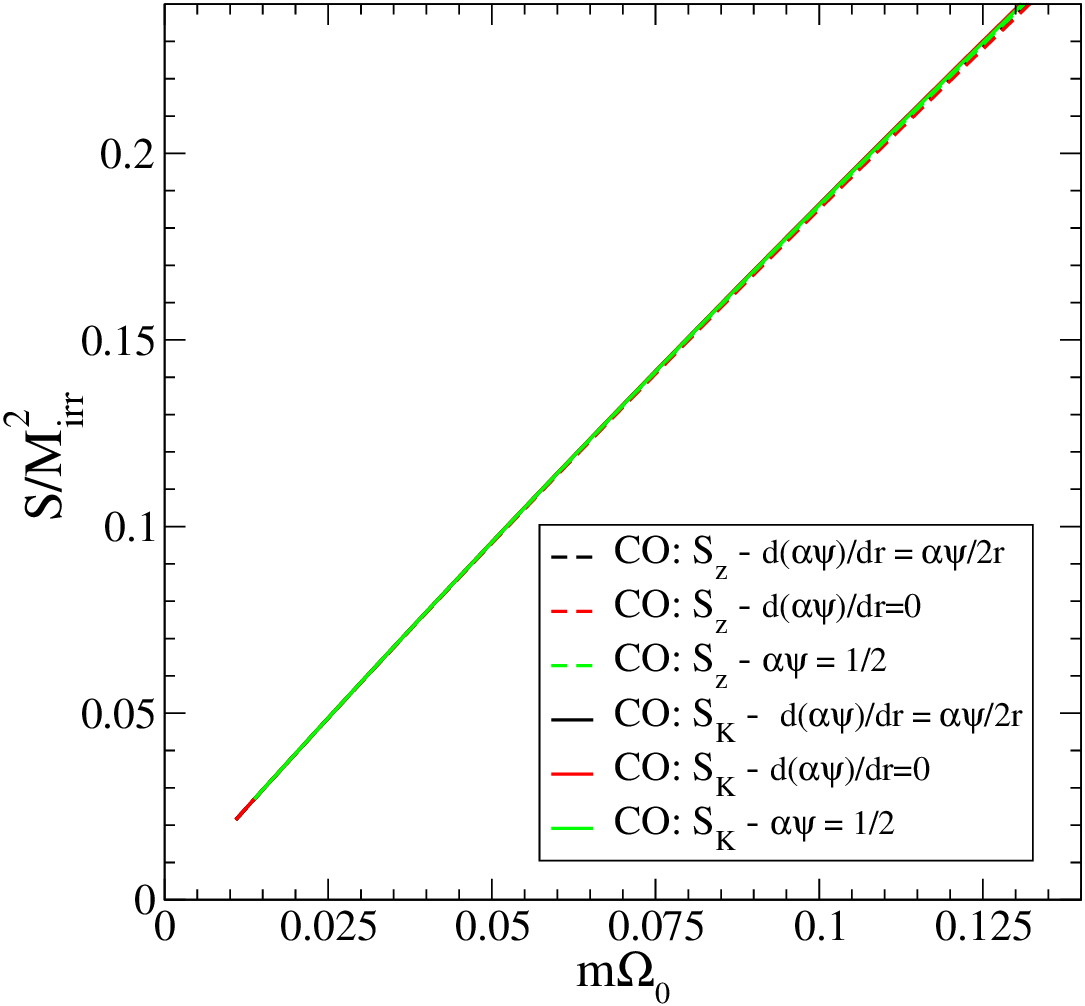}
\caption{\label{Fig:CoSpinKV} Quasi-local spin $S/M^2_{\rm irr}$ of a
black hole in a corotating equal-mass (CO) binary along the sequence of
circular orbits (parameterized by the orbital angular velocity
$m\Omega_0$).  $S_z$ and $S_K$ are defined in
Eq.~(\ref{eq:Spin-defs}). Results are given for sequences with three
different lapse boundary conditions.}
\end{figure}

We begin by first computing the spin of each black hole as a function
of the orbital angular velocity $\Omega_0$ along a sequence of circular
orbits that satisfy the Komar-mass condition.  In
Fig.~\ref{Fig:CoSpinKV}, we plot the quasi-local spin as computed by
Eq.~(\ref{eq:QL_spin}) for the three lapse boundary conditions listed
in Eqs.~(\ref{eq:Lapse-BCs}) and for the cases
where $\xi^i$ is defined via the flat-space Killing vectors or via an
approximate Killing vector as described in
Sec.~\ref{sec:comp-quasi-local}.  When computed using the
flat-space Killing vectors as define by
Eqs.~(\ref{eq:CKVs}), we note that $S_x$ and $S_y$
vanish to roundoff error.  We see that all cases show very close
agreement for the spin of the black holes.  Also, recall that we
expect our results to be largely independent of the choice of the
lapse boundary conditions and plot the results from different
lapse boundary conditions to confirm this conjecture.

While the various measures of the spin agree well, we have yet to
determine if the magnitude of the spin is correct.  To do so, we first
note that the spin of a Kerr black-hole, $S_{\rm Kerr}$, is given in
terms of its irreducible mass $M_{\rm irr}$ and the rotational angular 
frequency of its horizon, $\Omega_B$, by 
\begin{equation}\label{eq:Kerr_spin}
S_{\rm Kerr}(M_{\rm irr}, \Omega_B) =
      \frac{4 M_{\rm irr}^3\Omega_B}{\sqrt{1 - 4(M_{\rm irr}\Omega_B)^2}}.
\end{equation}
So, if we know the rotational angular frequency of the black hole's horizon, 
then we can check how well $S_z$ and $S_K$ satisfy Eq.~(\ref{eq:Kerr_spin}).  
Conversely, by inverting Eq.~(\ref{eq:Kerr_spin}) we can  estimate the angular
velocity $\Omega_B$ of the black hole in our numerical initial data from
the measured quasi-local spins.  It will be convenient to express this as
\begin{equation}\label{eq:Kerr_omegaratio}
  \frac{\Omega_B}{\Omega_0} = \frac{m}{M_{\rm irr}}\frac1{m\Omega_0}
    \frac{S/M^2_{\rm irr}}
	 {4\sqrt{1 + \frac14\left(S/M^2_{\rm irr}\right)^2}}.
\end{equation}

\begin{figure}
\includegraphics[width=\linewidth]{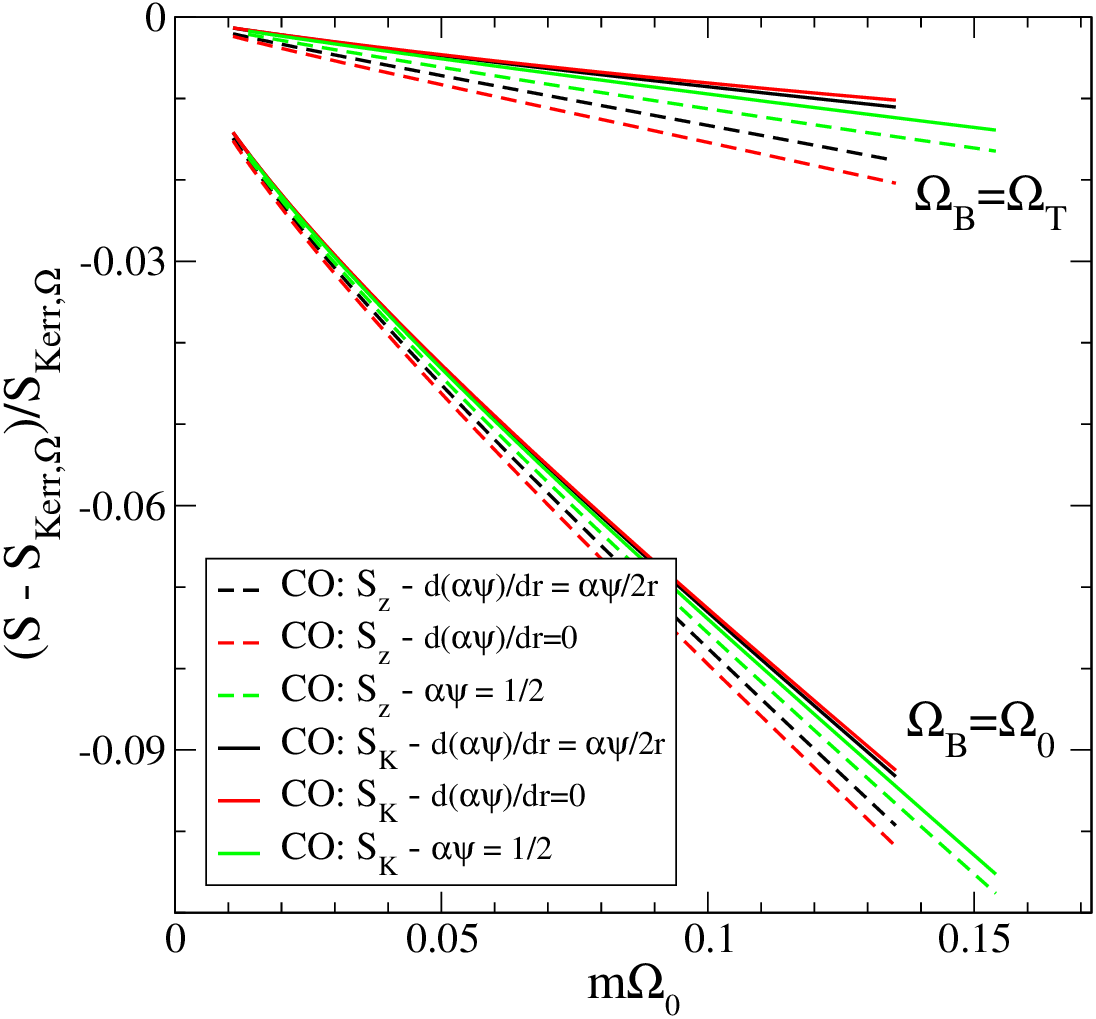}
\caption{\label{Fig:CoSpinErrKVLARF}Difference between the quasi-local
spin of a black hole in a corotating (CO) binary and the result of the
Kerr-formula Eq.~(\ref{eq:Kerr_spin}).  For the lower set of six lines
($\Omega_B=\Omega_0$), the rotation rate of the black hole is simply
taken to be equal to the orbital angular velocity.  For the upper set
of six lines ($\Omega_B=\Omega_T$), the rotation rate is taken to be
equal to that of the tidal field of the companion hole as measured in
the LARF.  Each set of six lines corresponds to the cases plotted in
Fig.~\ref{Fig:CoSpinKV}.  }
\end{figure}

In Ref.~\cite{cook-pfeiffer-2004a}, we assumed that the
rotational angular velocity of the black holes in a corotating binary 
was equal to the orbital angular velocity $\Omega_B=\Omega_0$.  We will 
refer to this assumption as the ``leading-order'' estimate.  The lower
six lines of Fig.~\ref{Fig:CoSpinErrKVLARF} display the relative error 
between the measured quasi-local spin and the spin of a Kerr black hole,
Eq.~(\ref{eq:Kerr_spin}), 
with the same irreducible mass and with $\Omega_B=\Omega_0$.  We see that the
quasi-local spin estimate derived from the approximate Killing vectors
is slightly better than the estimates based on the flat-space Killing
vectors.  Furthermore, the variation due to the lapse boundary
condition is smaller for the approximate Killing vector cases.  In all
cases, the relative error is smaller than about 8\% up to the ISCO
which occurs near $m\Omega_0\approx 0.1$, a quite large disagreement.

In order to make a better estimate of the error in the quasi-local
spin measurements, we need a better estimate of the correct spin
value.  To improve on this estimate we note that Alvi\cite{Alvi-2000}
has computed the leading order correction to the rotation rate $\Omega_T$
of the tidal field of a companion star as measured in the local asymptotic
rest frame (LARF)\cite{ThorneHartle-1985} of a black hole
\begin{equation}\label{eq:Alvi_LARF}
  \Omega_T = \Omega_0\left[1 - \eta\frac{m}{b} + 
     {\cal O}\left(\frac{m}{b}\right)^{3/2}\right],
\end{equation}
where $\eta\equiv\mu/m$ and $b$ is the separation of
the black holes in harmonic coordinates.  We can express this
in a gauge independent way using a post-Newtonian expansion
for $b/m$ obtained for circular orbits\cite{kidder95}
\begin{equation}\label{eq:PN_sep}
\frac{m}{b} = (m\Omega_0)^{2/3}\left[
     1 + (1 - \mbox{$\frac13$}\eta)(m\Omega_0)^{2/3} +
     {\cal O}(m\Omega_0)\right].
\end{equation}
Substituting, we find
\begin{align}\label{eq:PN_rotation}
  \frac{\Omega_T}{\Omega_0} &= 1 - \eta(m\Omega_0)^{2/3} 
    + \Lambda(m\Omega_0) \\ &\mbox{\hspace{0.25in}} 
    - \left[\eta(1 - \mbox{$\frac13$}\eta)-\Gamma\right](m\Omega_0)^{4/3}
    + {\cal O}(m\Omega_0)^{\frac53}, \nonumber
\end{align}
where $\Lambda$ and $\Gamma$ are functions of $\eta$ coming from the
unknown terms of order $(m/b)^{3/2}$ and $(m/b)^2$ respectively in
Eq.~(\ref{eq:Alvi_LARF}).  The leading order error term in
Eq.~(\ref{eq:PN_sep}), which includes spin-orbit coupling terms,
contributes to the term of order $(m\Omega_0)^2$ in
Eq.~(\ref{eq:PN_rotation}).  We now make the physical assumption that
$\Omega_T$ represents the angular velocity $\Omega_B$ at which a black
hole should rotate in its LARF in order to be in corotation.  This
assumption improves the agreement with the Kerr-formula dramatically,
as shown by the upper set of six lines in
Fig.~\ref{Fig:CoSpinErrKVLARF}.

\begin{figure}
\includegraphics[width=\linewidth]{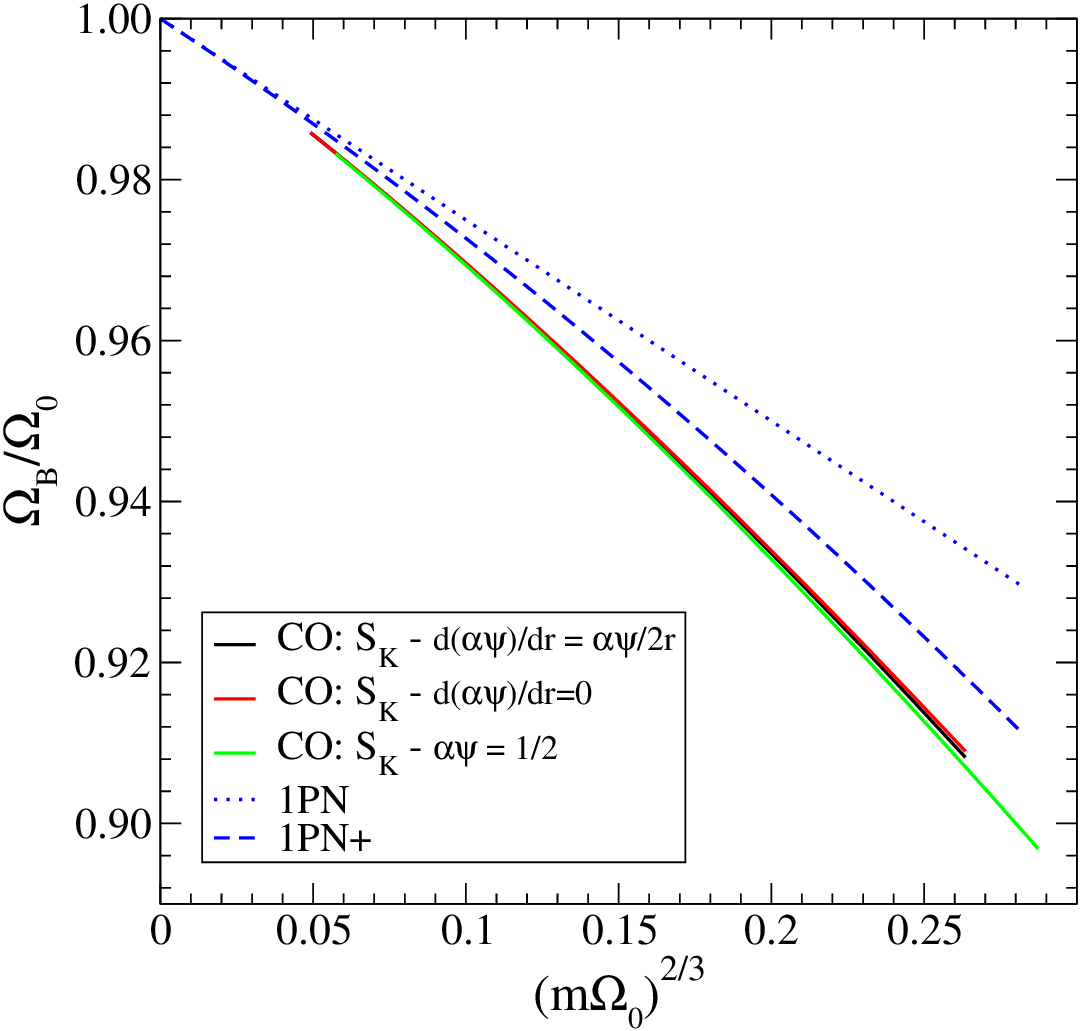}
\caption{\label{Fig:CoOmegaRatio} Ratio of the rotational angular
velocity $\Omega_B$ of a black hole, as determined from the
quasi-local spin $S_K$, to the orbital angular velocity of the binary
along the corotating (CO) sequence.  The dotted line (labeled {\tt
1PN}) represents the leading order correction to the analytic estimate
of the ratio.  The dashed line (labeled {\tt 1PN+}) shows the higher
order correction of Eq.~(\ref{eq:PN_rotation}) with $\Lambda=\Gamma=0$.
Numerical results are given for three different lapse boundary conditions.
}
\end{figure}

Figure~\ref{Fig:CoOmegaRatio} shows, from a different perspective, the
improvement obtained by using Alvi's result.  From
Eq.~(\ref{eq:Kerr_omegaratio}), we obtain the expected value of
$\Omega_B/\Omega_0$ from the initial data, {\em assuming} the
Kerr-formula is exactly satisfied.  In Fig.~\ref{Fig:CoOmegaRatio},
we show this expected value for $\Omega_B/\Omega_0$ for each of the
three lapse boundary conditions and compare these to
$\Omega_T/\Omega_0$ from Eq.~(\ref{eq:PN_rotation}).  The dotted line
labeled {\tt 1PN} includes only the leading-order correction to
$\Omega_T/\Omega_0$.  Since we are plotting the ratio against
$(m\Omega_0)^{2/3}$, this line is linear.  The dashed line labeled
{\tt 1PN+} shows the full analytic estimate assuming the unknown
coefficients $\Lambda=\Gamma=0$.  For our equal-mass binaries, $\eta = 1/4$.

We see that the analytic estimate for the rotation rate is in
reasonably good agreement with the numerical results.  In fact, the
difference between them is nearly linear in $m\Omega_0$ and is well
fit by setting $\Lambda\sim-0.085$(for the equal-mass case of
$\eta=1/4$ and $\Gamma=0$).  It would be quite interesting to
determine the next term in the rate of rotation of the tidal field
given in Eq.~(\ref{eq:Alvi_LARF}) since this term would fix $\Lambda$.
We should note, however, that we cannot expect our estimates of the
spin to agree perfectly with the higher-order analytic estimates of
the black-hole rotation rate for corotation.  There are two reasons
for this.  One is that our initial data is conformally flat.  In
Ref.\cite{cook-pfeiffer-2004a}, we examined the case of a single
rotating, conformally-flat black hole.  Maximum rotation in a
corotating binary occurs at ISCO where the rotational angular velocity
has a value of $M_{\rm irr}\Omega_B\sim0.05$.  At this rate of spin,
we expect the approximation of conformal flatness to introduce an
error of about 0.02\%.  The second source of error is the quasi-local
estimate of the spin itself which has some {\em a priori} unknown
level of uncertainty.  We show a rough indication of the current level
of uncertainty in Fig.~\ref{Fig:CoSpinErrKVLARF}.  In the top six
lines, we again plot the relative error of the quasi-local spins, but
now plotting the error relative to the Kerr spin based on the tidal
rotation rate given by Eq.~(\ref{eq:PN_rotation}) with $\Lambda=\Gamma=0$

We see that at ISCO ($m\Omega_0\sim0.1$), the relative errors for the
spins computed using an approximate Killing vector are all less than
1\%, but significantly larger than the 0.02\% error caused by the
assumption of conformal flatness.  Unfortunately, we cannot determine
whether this remaining error is due to the inherent uncertainty of our
quasi-local measure of the spin or to the unknown value of $\Lambda$ 
(and higher-order terms).
The only thing that we can conclude is that the uncertainty in the
spin based on comparing the two different approaches to compute the
spin is of the order of 1\%, and is about 0.5\% based on the
uncertainty introduced by the different lapse boundary conditions.
Both are of the same order as the total error which is remarkably
small and gives us considerable confidence in our quasi-local spin
measure.

\section{Non-Spinning Binaries}
\label{sec:non-spin-binary}

In order to model non-spinning binaries, we must choose an appropriate
non-vanishing excision boundary condition on $\parShift^i$.  Since our
models are set up so that the orbital angular momentum is pointed in
the positive $z$ direction, we choose the boundary condition on
$\parShift^i$ via Eq.~(\ref{eq:parShift_BC}) and set $\xi^i =
\xi^i_{(z)}$, the flat-space Killing vector of Eq.~(\ref{eq:CKV_z}).
This choice for $\xi^i$ is appropriate because we have a spherical
excision surface in a flat conformal geometry, so $\xi^i_{(z)}$ is a
conformal Killing vector of the physical induced metric on the
excision surface.  The only freedom remaining in the boundary
condition is the magnitude of $\Omega_r$.

\subsection{Leading-order approximation: \boldmath$\Omega_r=\Omega_0$}
\label{sec:lead-order-NS}

In Ref.~\cite{cook-pfeiffer-2004a}, it was argued that, at least to
leading order, the correct choice for non-spinning black holes was to
choose $\Omega_r$ to be equal to the orbital angular velocity,
$\Omega_r = \Omega_0$.  The numerical results were in good agreement
with post-Newtonian estimates for sequences of equal-mass non-spinning
binaries in circular orbits and with post-Newtonian estimates for the
location of the ISCO.  In this section, we revisit this argument with
the new diagnostic tools presented earlier in this paper, in
particular, effective potential plots and quasi-local measures of the
spin.

\begin{figure}
\includegraphics[width=\linewidth]{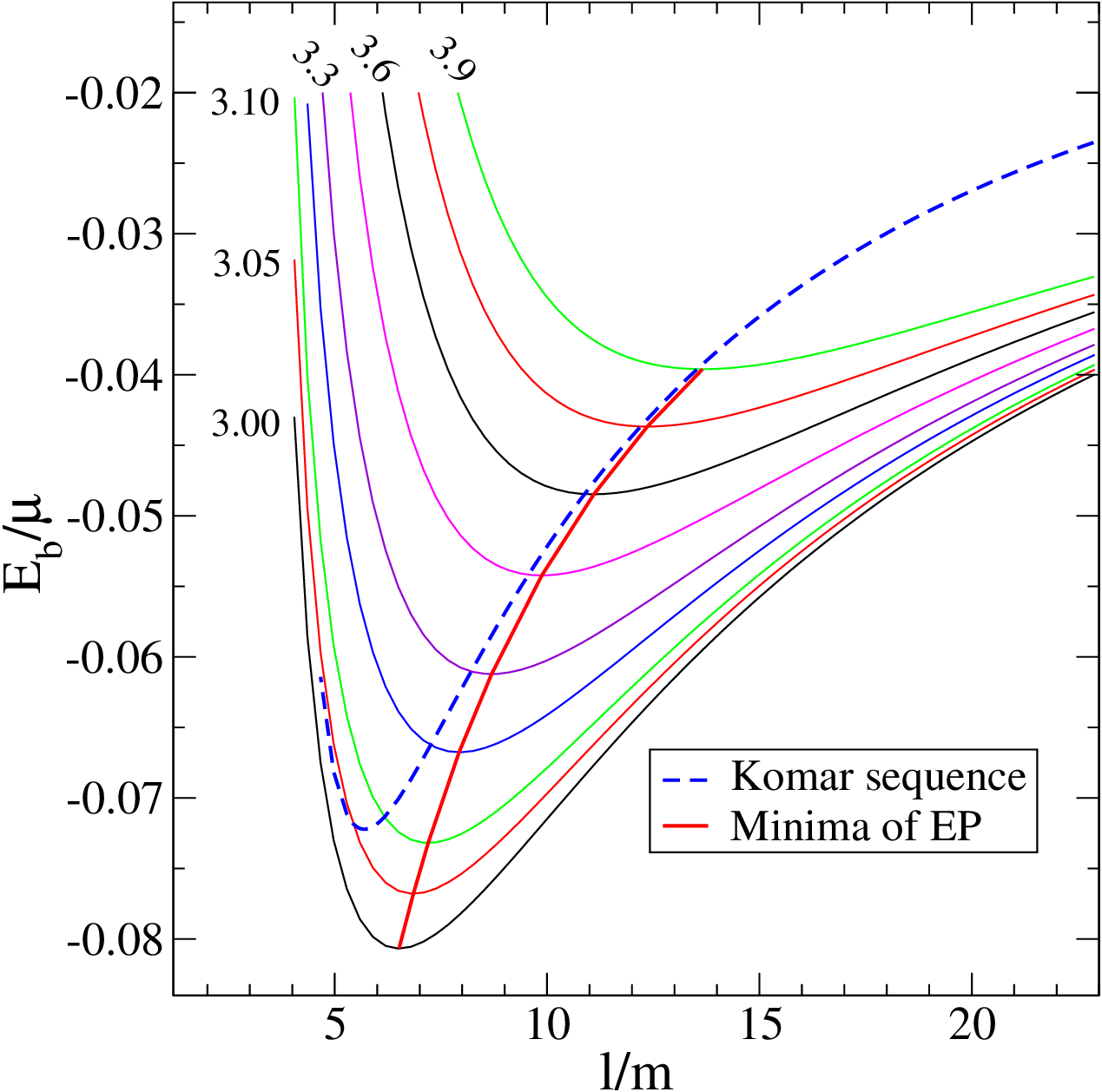}
\caption{\label{Fig:EP_irrot} EP curves $E_b/\mu$ for equal-mass
``leading-order'' non-spinning binaries ($\Omega_r\!=\!\Omega_0$)
plotted vs.  separation $\ell/m$.  These curves are labeled by the
value of $J/\mu m$ along each curve.  Also plotted are the line
connecting the minima of the EP-curves, as well as the sequence of
circular orbits as determined by the Komar-mass ansatz.  The
Komar-mass ansatz and the effective potential method clearly disagree.}
\end{figure}

Figure~\ref{Fig:EP_irrot} displays several EP curves for the
leading-order (i.e. $\Omega_r=\Omega_0$) non-spinning binaries where
the total angular momentum is held fixed.  It is immediately clear
that the sequence of circular orbits determined by the Komar-mass
condition does not intersect the minima of the EP curves.
Furthermore, there does not appear to be an inflection point in the EP
curves anywhere near the minimum of the Komar sequence of circular
orbits.  This clearly indicates that a problem of some kind exists.

\begin{figure}
\includegraphics[width=\linewidth]{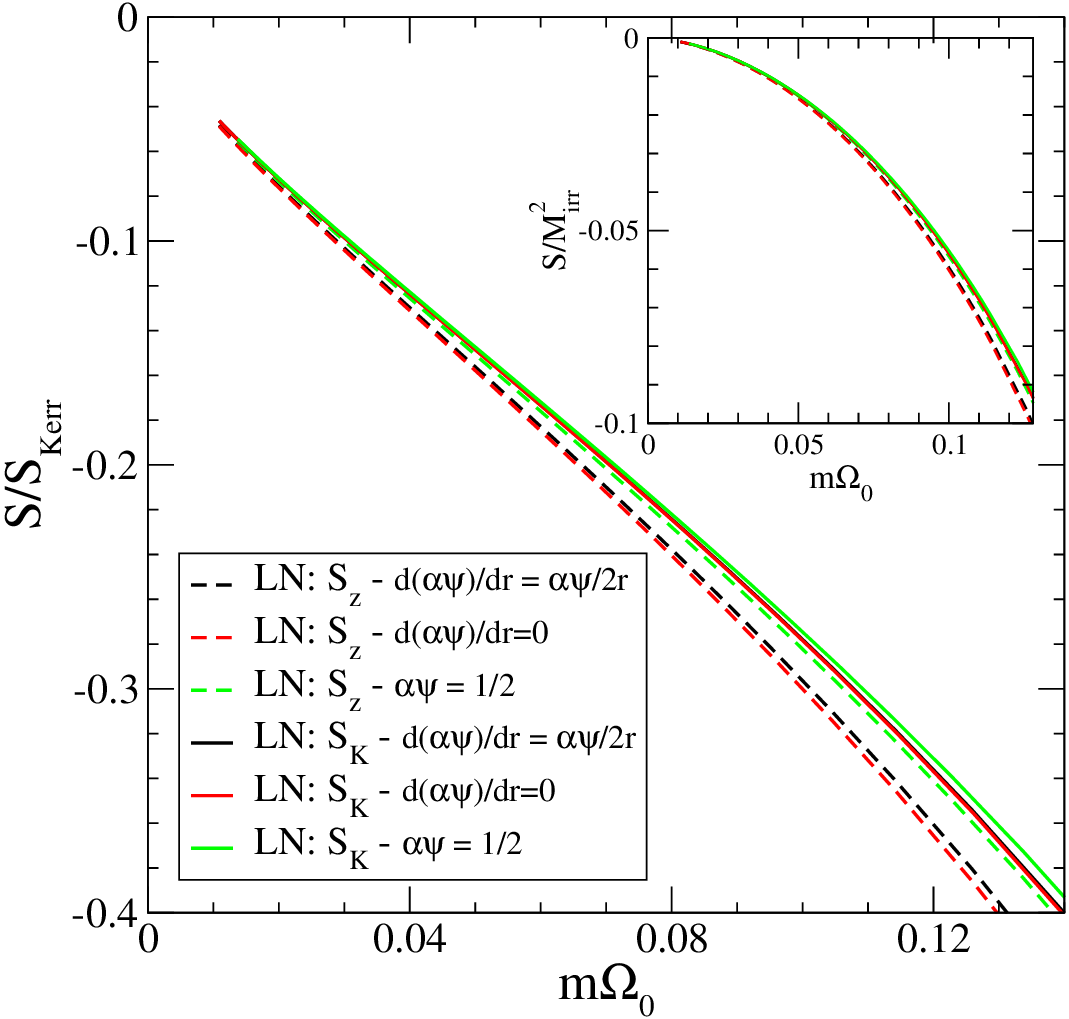}
\caption{\label{Fig:IrSpinErrKV} Quasi-local spin of the ``leading
order'' non-spinning (LN) binaries (as shown in Fig.~\ref{Fig:EP_irrot})
plotted vs. orbital angular velocity.  The inset displays the
dimensionless spin value $S/M^2_{\rm irr}$.  The main figure displays
the magnitude of the spin relative to the spin of a Kerr black hole
with $\Omega_B=\Omega_0$.  The residual spin is quite large.}
\end{figure}

The source of the problem is made clear by examining the quasi-local
spins of black holes in our leading-order non-spinning models.  The
inset of Fig.~\ref{Fig:IrSpinErrKV} shows the magnitude of the spin on
one of the black holes in the leading-order non-spinning equal-mass
binary as a function of the orbital angular velocity.  As in the
corotating case, the $x$ and $y$ components of the spin vanish to
roundoff error and we see that the $z$ component of the spin computed
using flat-space Killing vectors agrees well with the magnitude of the
spin computed using an approximate Killing vector.  But, it is clear
that the magnitude of the spin is not zero.  To get a better
understanding of whether or not this magnitude of spin is small,
Fig.~\ref{Fig:IrSpinErrKV} also compares this spin with the spin of a
Kerr black hole having a rotational angular velocity equal to the
binary's orbital angular velocity.  Essentially, this is a measure of
the magnitude of the spin relative to the corresponding corotating
case.

We see that the spin for the leading-order non-spinning equal-mass
binaries can be as large as 30\% of the corotation spin at ISCO.  This
is clearly a significant deviation from the desired non-spinning
configuration we wish to model.

Another indication that there is an error in the leading-order
non-spinning equal-mass binaries is found in examining how well the
sequence of leading-order non-spinning circular orbits agrees with the
thermodynamic identity of Eq.~(\ref{eq:thermo_ident0}).  In
Table~\ref{tab:thermo_irr_mass}, we show the results for sequences of
circular orbits constructed using the Komar-mass condition.  The last
column shows the results for leading-order non-spinning binaries.  We
find the disturbing result that the irreducible mass decreases as the
binary evolves to smaller radii.  More importantly, comparing this
variation to that seen in the second column, we find that the size of
the variation in $M_{\rm irr}$ for the leading-order non-spinning case
is roughly 20 times larger than that seen in the corotating case (see
also Fig.~9 of Ref.~\cite{cook-pfeiffer-2004a}).  Of course, the
thermodynamic identity was derived for corotating binaries, not
non-spinning binaries so we shouldn't be surprised that it is not
satisfied in this case, and by itself, this failure is not too
worrisome.  However, the combined evidence in this section, in
particular Figures \ref{Fig:EP_irrot} and \ref{Fig:IrSpinErrKV}, makes it
clear that the leading order approximation $\Omega_r=\Omega_0$ is not
satisfactory. Since we can now evaluate the quasi-local spin, we no
longer have to choose $\Omega_r$ by some {\em ad hoc} prescription --
instead, we can simply choose it such that the quasi-local spin
vanishes.  This approach will be explored in the following section.

\subsection{Correct approach for non-spinning binaries}
\label{sec:correct-approach}

To correctly model non-spinning black-hole binaries, we need to choose
$\Omega_r$ so that the quasi-local measure of the spin vanishes,
$S_K=0$.  This will involve root-finding, i.e., the constraints will
have to be solved for different values $\Omega_r$ until the solution
satisfies $S_K=0$.  Since the correct value of $\Omega_r$ for
non-spinning black holes will be close to $\Omega_0$, we define a
rotation fraction $f_r$ as
\begin{equation}\label{eq:rot_fraction}
  \Omega_r = f_r\Omega_0.
\end{equation}
While the definition of $f_r$ mirrors Eq.~(\ref{eq:PN_rotation}) from
our comparison with post-Newtonian results, we note that those
post-Newtonian results do {\em not} enter the construction of
non-spinning binary black hole initial data.  In practice, we
determine $f_r$ so that the quasi-local spin based on the approximate
Killing vector vanishes to about one part in $10^8$ or better.

\begin{figure}
\includegraphics[width=\linewidth]{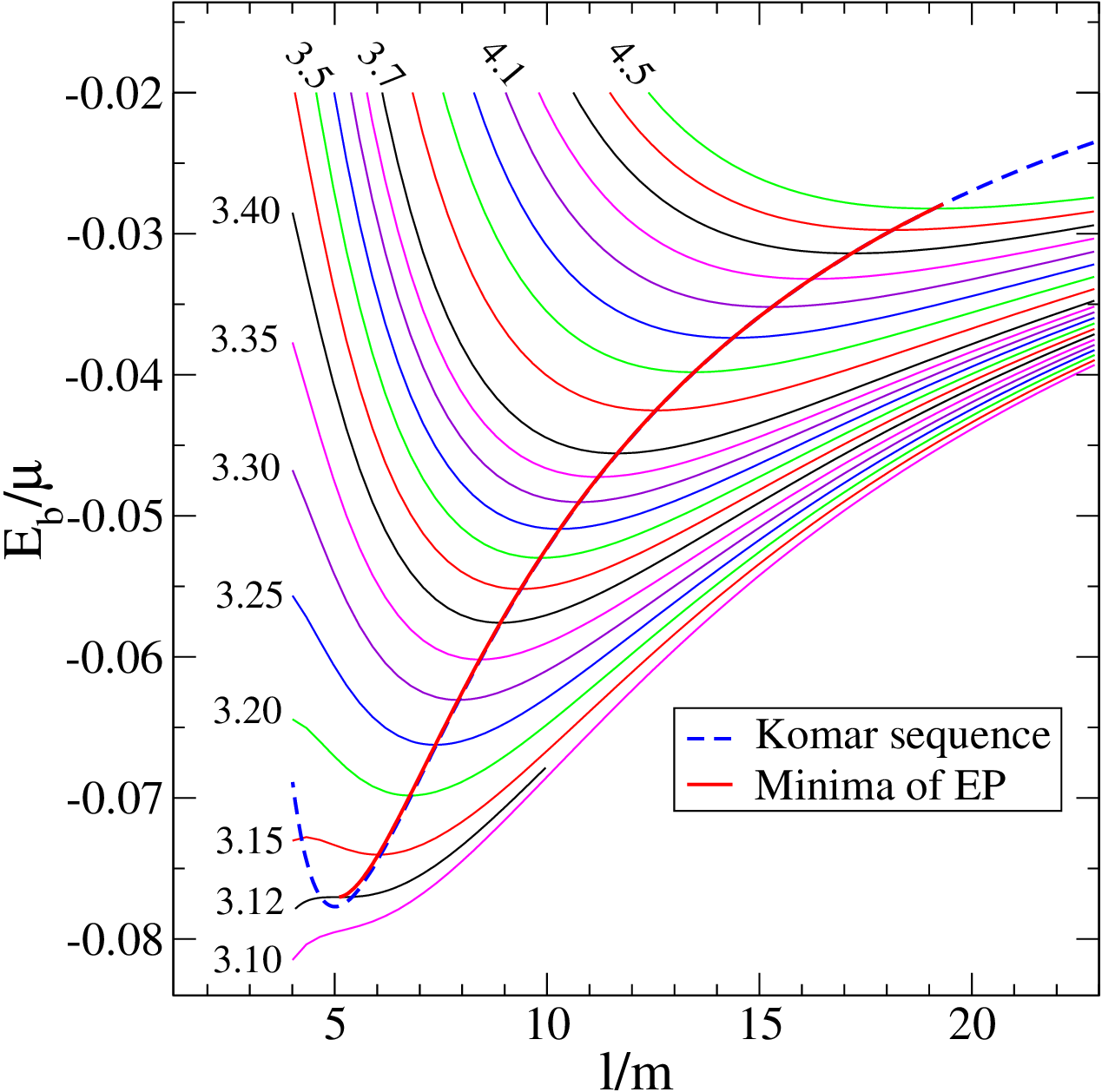}
\caption{\label{Fig:EP_tirot} EP curves $E_b/\mu$ for ``true''
non-spinning black holes vs. separation $\ell/m$.  The curves are
labeled by the orbital angular momentum $J/\mu m$ which is kept
constant along each curve.  The thick red line connecting the minima
of the EP-curves represents circular orbits, and terminates at the
innermost stable circular orbit at the inflection point in the
EP-curve at $J/\mu{m}=3.12$.  Also plotted as a dashed blue line is
the sequence of circular orbits as determined by the Komar-mass
ansatz.}
\end{figure}

In Fig.~\ref{Fig:EP_tirot} we first plot the EP curves for ``true''
non-spinning equal-mass black-hole binaries.  The solid line passing
through the minima of the EP curves is the sequence of circular orbits
as defined by the EP method.  This sequence terminates at small
$\ell/m$ at the ISCO which is defined by the occurrence of an inflection
point in an EP curve.  Finally, plotted as a dashed line in this figure
is the sequence of circular orbits as defined by the Komar-mass
condition.  It is immediately apparent that the EP curves have the
correct qualitative behavior and that the Komar sequence of circular
orbits, drawn as a dashed line, passes very close to the minima of the
EP curves everywhere except near the ISCO.

\begin{figure}
\includegraphics[width=\linewidth]{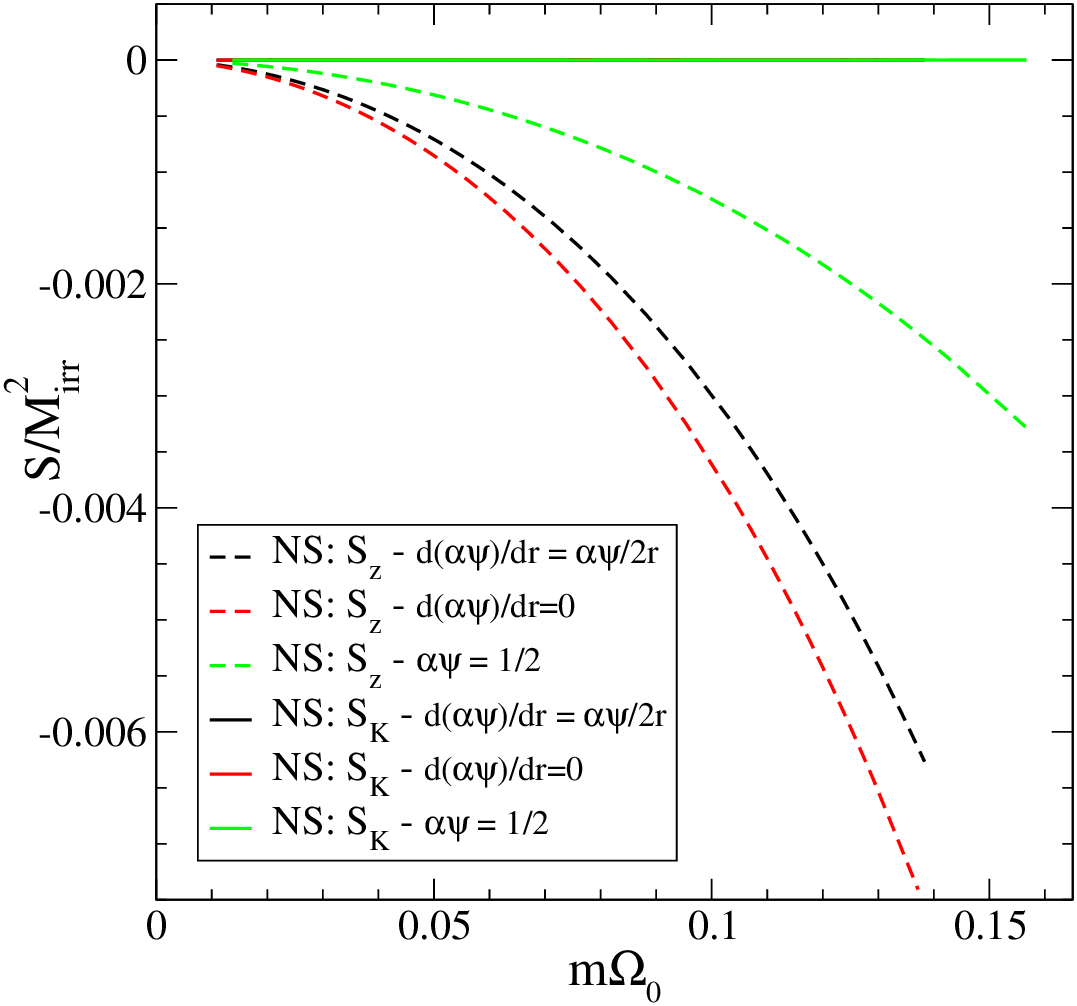}
\caption{\label{Fig:TiSpinErr} Quasi-local spin for a ``true''
non-spinning (NS) black hole in an equal-mass binary along the sequence of
circular orbits, parametrized by the orbital angular velocity.  The
same six cases are plotted as in the inset of
Fig.~(\ref{Fig:IrSpinErrKV}).  Note that here $S_K=0$ by construction.}
\end{figure}

In Fig.~\ref{Fig:TiSpinErr}, we plot the value of the quasi-local spin
of one black hole in an equal-mass binary as a function of the orbital
angular velocity.  The sequences are circular orbits defined by the
Komar-mass condition and we include plots of the three different lapse
boundary conditions for both definitions of the spin.  By construction, 
the quasi-local spin defined by the
approximate Killing vector is zero, $S_K=0$.  However, the quasi-local
spin defined by the flat-space Killing vector ($S_z$) is not
necessarily zero and is a rough measure of the uncertainty in our
definition of the spin.  For separations larger than ISCO, we see that,
in terms of $S_z$, the spin does not exceed 0.4\% of maximal rotation.

\begin{figure}
\includegraphics[width=\linewidth]{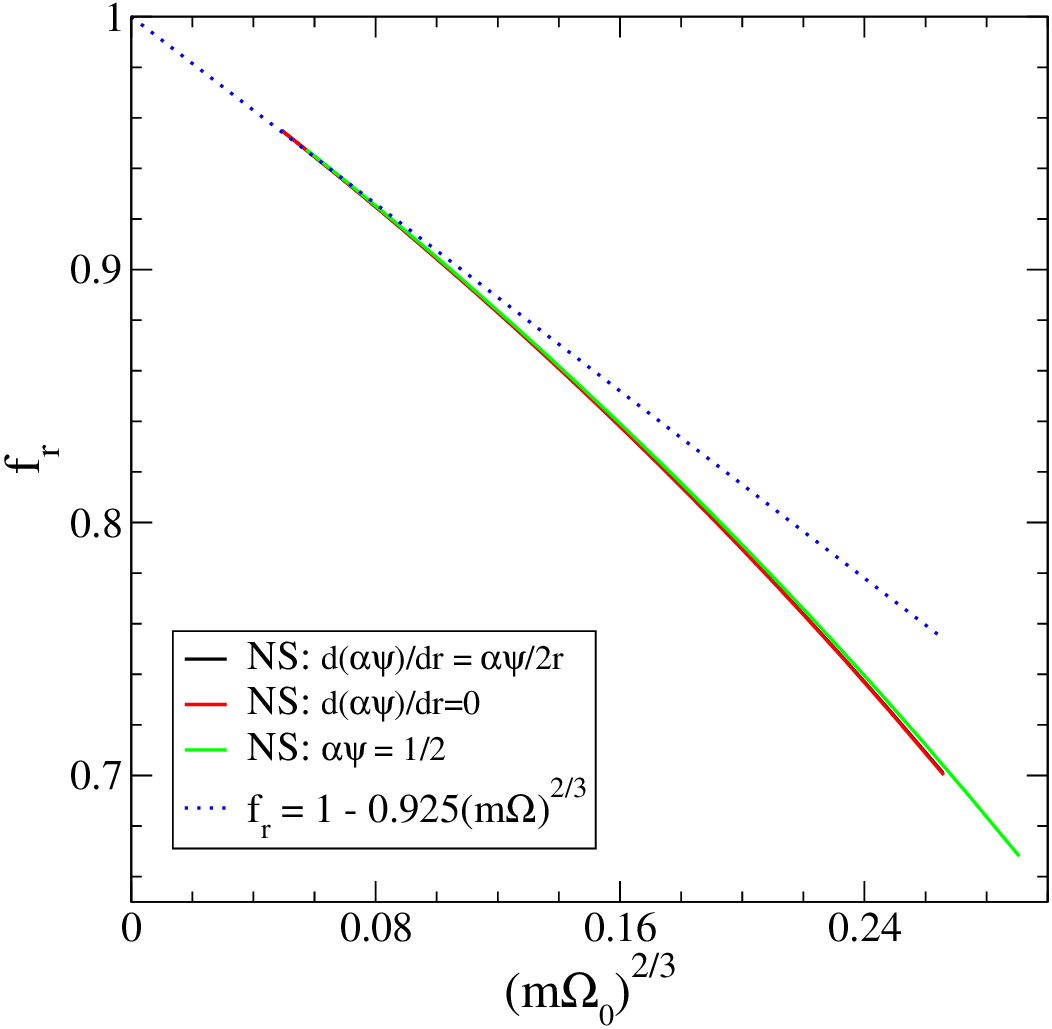}
\caption{\label{Fig:TiOmegaRatio} Rotation parameter $f_r =
\Omega_r/\Omega_0$ along the sequence of non-spinning (NS) black
holes, parametrized by the $2/3$-power of the orbital angular
velocity.  The dotted line plots the fit to $f_r$ through the term of
order $(m\Omega_0)^{2/3}$.}
\end{figure}

In Fig.~\ref{Fig:TiOmegaRatio}, we plot the value of the rotation
fraction $f_r=\Omega_r/\Omega_0$ as a function of
$(m\Omega_0)^{2/3}$.  To leading order, we expected $f_r=1$ and that
to model non-rotating black holes, we needed to set the excision
boundary conditions on $\parShift^i$ so as to ``unrotate'' the black
hole from the case of corotation when $\parShift^i=0$.  We might then
expect that at higher order we would find $f_r = \Omega_B/\Omega_0 \approx
\Omega_T/\Omega_0$.  However, comparing $f_r$ in
Fig.~\ref{Fig:TiOmegaRatio} to $\Omega_B/\Omega_0$ in
Fig.~\ref{Fig:CoOmegaRatio} we see that $f_r$ is significantly
smaller.  In terms of the previous physical interpretation of the
excision boundary conditions for non-spinning holes, this implies that
we need to ``unrotate'' the black hole by less than the corotation
rate.  Currently, we do not have a theoretical interpretation for the
value of $f_r$ seen in Fig.~\ref{Fig:TiOmegaRatio}.  If we assume that
an expansion in terms of $m\Omega_0$ takes the same functional form
as $\Omega_T/\Omega_0$ in Eq.~(\ref{eq:PN_rotation}), we find that
$f_r$ is well fit by
\begin{equation}\label{eq:NSratio}
  f_r = 1 - 0.925(m\Omega_0)^{2/3} 
    + 0.36(m\Omega_0) - 1.4(m\Omega_0)^{4/3}.
\end{equation}
We note, however, that $f_r$ can also be well fit by a function that
includes a term of order $(m\Omega_0)^{1/3}$.

\subsection{Results for non-spinning binaries}
\label{sec:results-non-spinning}

\begin{figure}
\includegraphics[width=\linewidth]{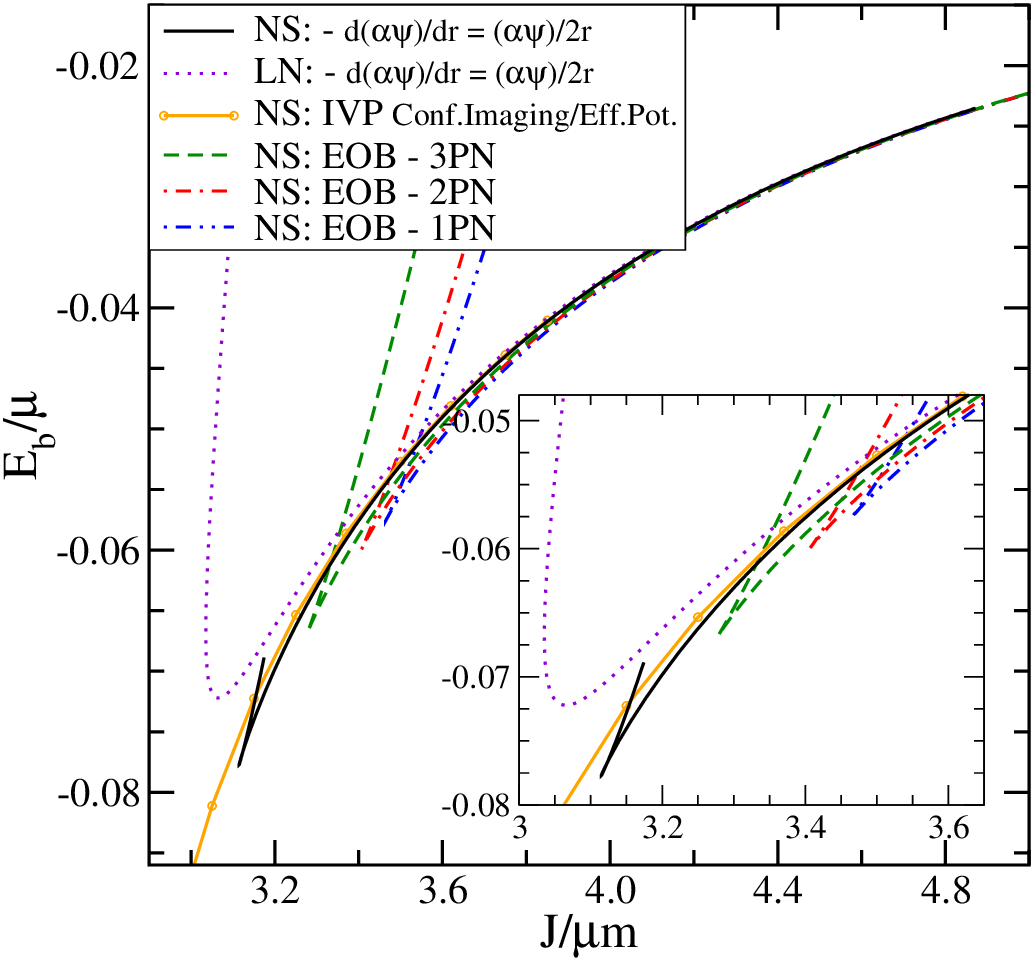}
\caption{Binding energy vs. total angular momentum along the sequence
 of {\em non-spinning} (NS) equal-mass black holes (as defined by the
 Komar-mass condition).  For comparison, the leading-order
 non-spinning (LN) results from our earlier
 work~\cite{cook-pfeiffer-2004a} are included, as well as results from
 Refs.~\cite{cook94e,Damour-etal-2002}.}
\label{fig:Eb_J_TICMP}
\end{figure}

\begin{figure}
\includegraphics[width=\linewidth]{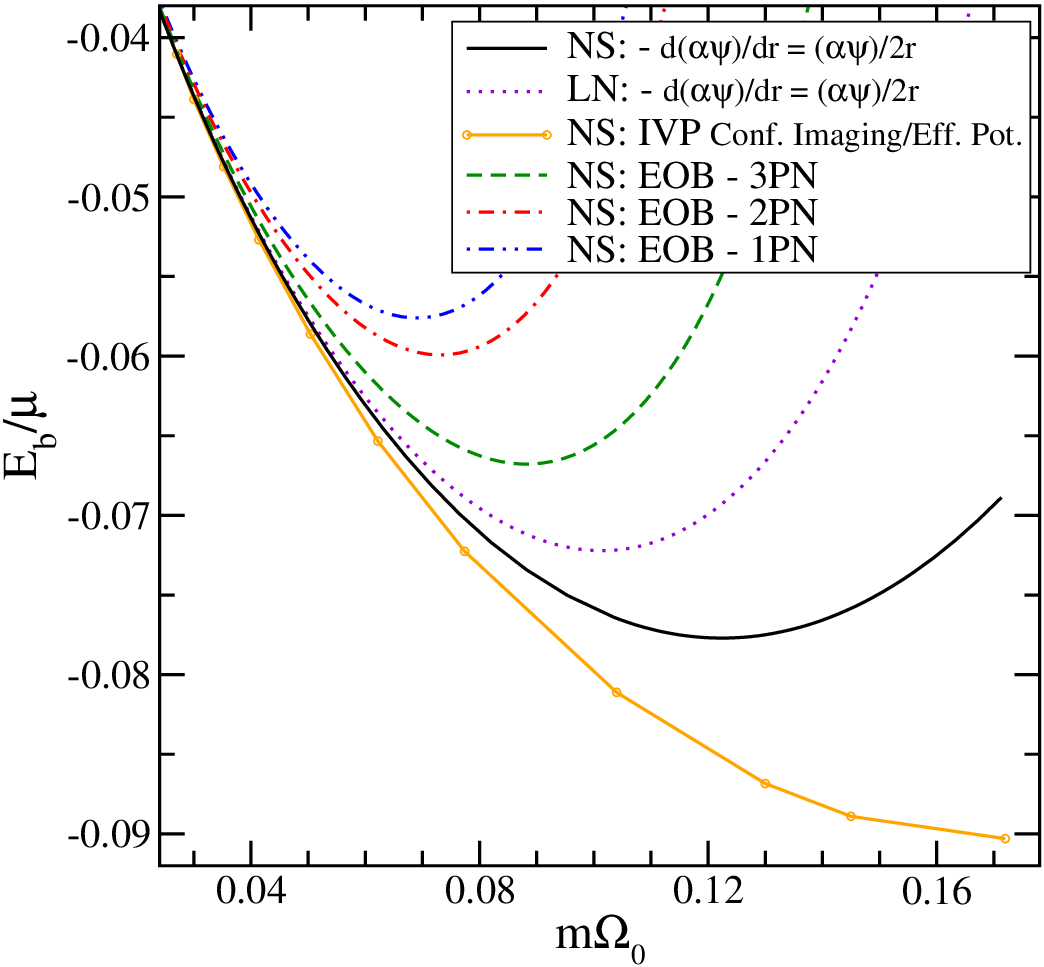}
\caption{Binding energy vs. orbital frequency along the
sequence of {\em non-spinning} (NS) equal-mass black holes (as defined by
the Komar-mass condition).  Lines are labeled as in Fig.~\ref{fig:Eb_J_TICMP}.}
\label{fig:Eb_O_TICMP}
\end{figure}

\begin{figure}
\includegraphics[width=\linewidth]{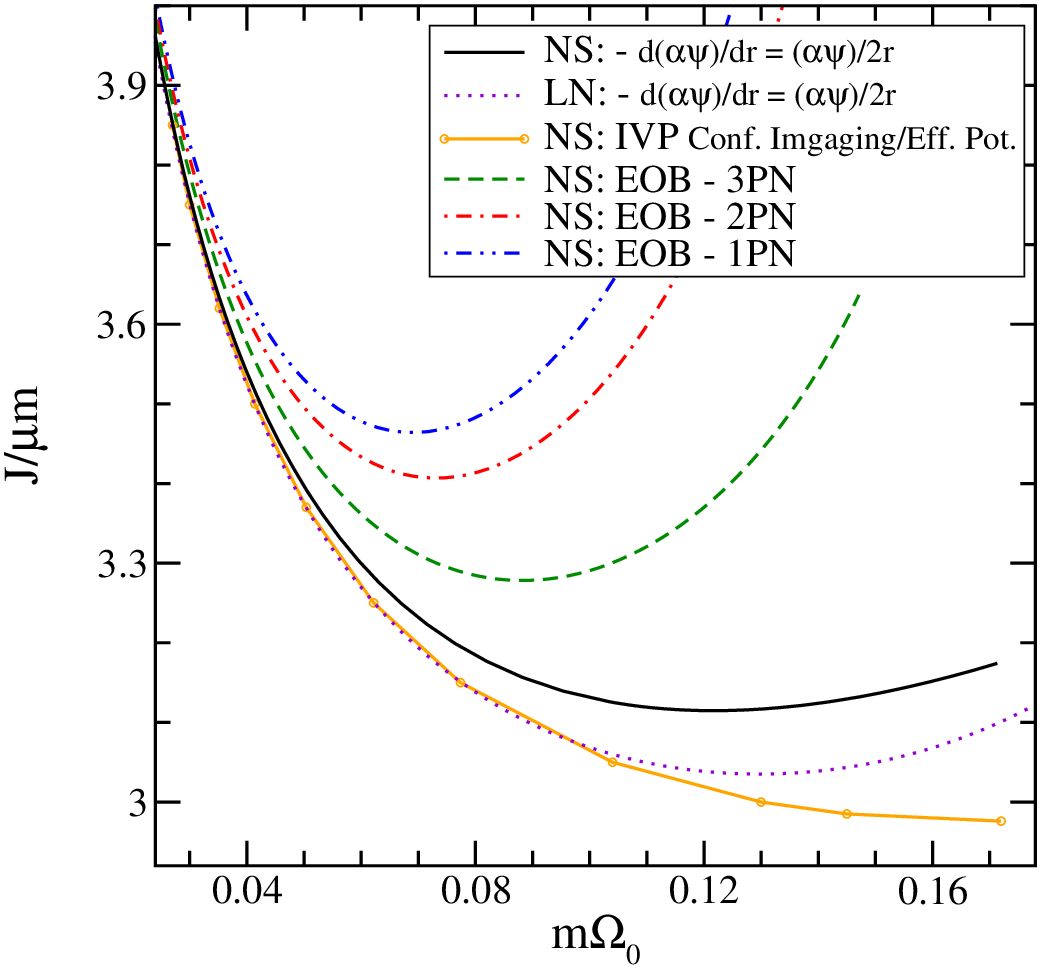}
\caption{Total angular momentum vs. orbital frequency along the sequence
 of {\em non-spinning} (NS) equal-mass black holes (as defined by the
 Komar-mass condition).  Lines are labeled as in Fig.~\ref{fig:Eb_J_TICMP}.}
\label{fig:J_O_TICMP}
\end{figure}

In Ref.~\cite{cook-pfeiffer-2004a}, we carefully examined the case of
equal-mass non-spinning binaries, but the initial-data sets were
constructed using the method for defining non-spinning black holes
that is correct only to leading order.  Because our improved approach
does yield a different spin for the resulting black holes, we
reexamine the physical content of these configurations.  In the
previous section, we computed the sequence of circular orbits for
truly non-spinning binaries (cf. Fig.~\ref{Fig:EP_tirot}).  In
Figs.~\ref{fig:Eb_J_TICMP}--\ref{fig:J_O_TICMP}, we plot parameters
along the sequence of non-spinning binaries defined using the
Komar-mass condition.  In each case, we compare our initial-data
sequences to the effective one body (EOB) post-Newtonian
results\cite{Damour-etal-2002,cook-pfeiffer-2004a} at first, second,
and third post-Newtonian order, to the results of an earlier
initial-data method\cite{cook94e}, and to the results based on the
leading-order method of defining non-spinning black holes.  While the
difference between our improved numerical results and those based on
the leading-order method are not dramatic, they do become significant
at small separations.

The most remarkable change produced by the improved method is seen in
Fig.~\ref{fig:Eb_J_TICMP} which plots the dimensionless binding energy
as a function of the dimensionless total angular momentum.  Using the
leading-order method, the sequence did not approximate a cusp at the
ISCO as we would have expected.  However the improved data clearly
approximates a cusp.  In Ref.\cite{cook-pfeiffer-2004a}, we pointed
out that we did not understand why our non-spinning data lacked this
feature.  We now understand that the approximate cusp is a necessary
feature of a sequence that is in good agreement with the EP method.
This can be seen by looking at Figs.~\ref{Fig:EP_irrot} and
\ref{Fig:EP_tirot} and considering the behavior near the inflection
point of Fig.~\ref{Fig:EP_tirot} that defines the ISCO for the EP
method.  In fact, a sequence of circular orbits defined by the EP
method necessarily has the minima in $E_b/\mu$ and $J/\mu{m}$ coincide,
resulting in an exact cusp in the sequence at ISCO.

We can also reexamine how well the improved non-spinning data agrees
with the thermodynamic identity of Eq.~(\ref{eq:thermo_ident0}).  The
third column of Table~\ref{tab:thermo_irr_mass} shows $M_{\rm irr}$
along a sequence of non-spinning equal-mass binaries in circular orbit
constructed using the Komar-mass condition.  Comparing to the results
of the second column for the corresponding case of corotation, we see
that the new approach for defining non-spinning binaries yields
results that are comparable in magnitude.  Also, the variation in
$M_{\rm irr}$ is much smaller than those seen in leading-order
non-spinning data of the last column.

\begin{figure}
\includegraphics[width=\linewidth]{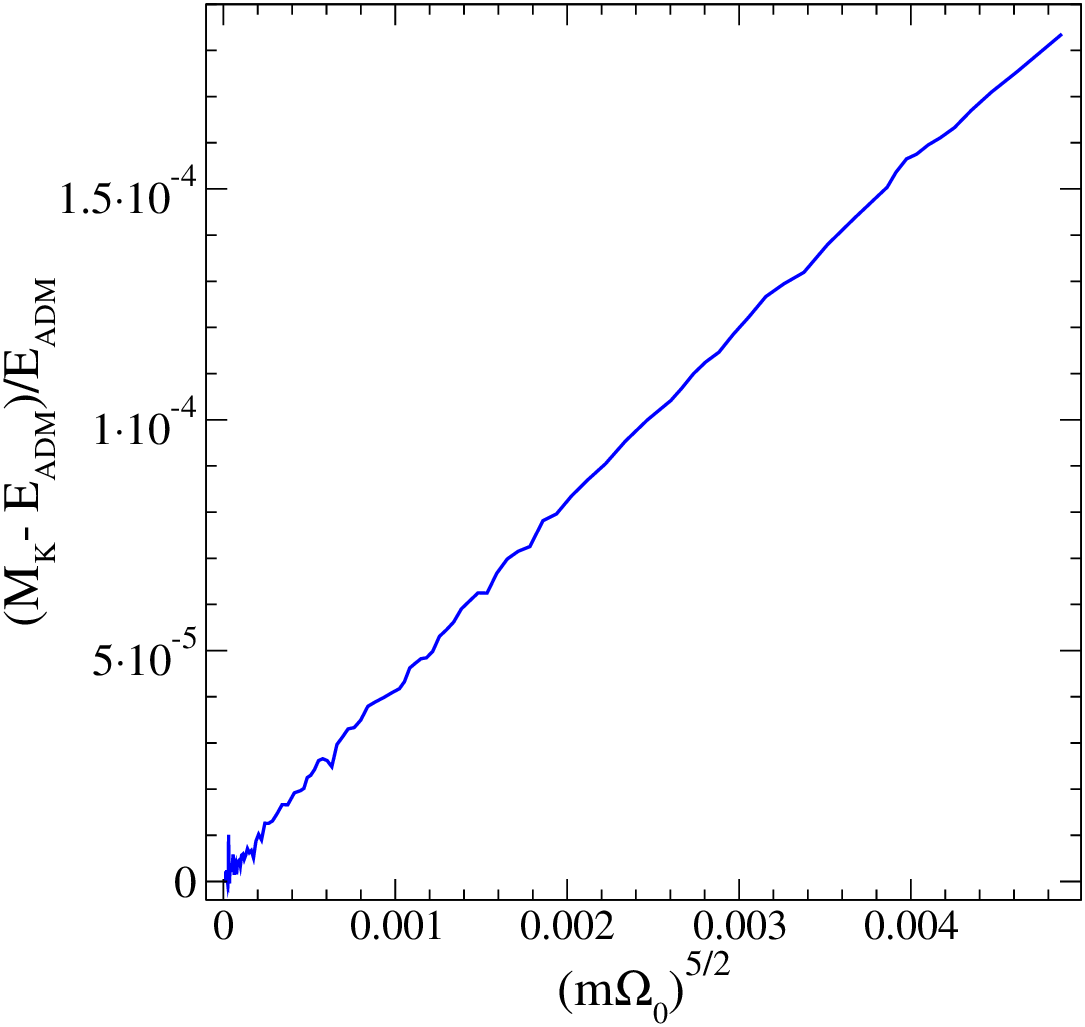}
\caption{\label{Fig:KomarErr_tirot} Violation of the Komar-mass
condition when the effective potential method is used to determine the
sequence of circular orbits. Here, {\em non-spinning} (NS)
equal-mass binaries are considered, and the sequence is parametrized
by the orbital angular momentum.  }
\end{figure}

Finally, it is interesting to reconsider the sequence of non-spinning
circular orbits constructed by the EP method.  Recalling that an exact
stationary solution of Einstein's equations for a black-hole binary
can only be found if the black holes are in corotation, we should
examine again how well the Komar-mass condition and the EP method agree
in their predictions of circular orbits.  From
Fig.~\ref{Fig:EP_tirot}, we see that the sequences of circular orbits
defined by the two methods nearly coincide except for the regime near
the ISCO.  For a more quantitative comparison, we can again examine
the relative error in the Komar-mass condition.
Figure~\ref{Fig:KomarErr_tirot} displays this error for the case of
non-spinning equal-mass binaries plotted against the dimensionless
orbital angular velocity.  The magnitude of the relative error is
comparable to that of the corotating case plotted in
Fig.~\ref{Fig:KomarErr_corot}.

We have also examined how well the sequence of non-spinning circular
orbits constructed by the EP method agrees with the thermodynamic
identity of Eq.~(\ref{eq:thermo_ident0}).  As with the corotating
case, we use the freedom to define the fundamental length scale along
the sequence to enforce Eq.~(\ref{eq:thermo_identity}).  Then the
deviation in $M_{\rm irr}$ along the sequence is a measure of how well
the thermodynamic identity is satisfied.  As with the case of
corotation, we find that the variation in the mass is smaller along
the EP sequence than in the Komar-mass sequence.  However, the
variations are {\em not} as small as seen in the corotating EP
sequence.  While the variation in the corotating EP sequence were
consistent with truncation error, the variations in the non-spinning
EP sequence are consistently larger by a factor of about five and the
increase in the mass appears to be systematic and significant.

This result should not be surprising since the thermodynamic identity
(\ref{eq:thermo_ident0}) we are testing has only been defined for the
case of corotating binaries.  What is remarkable is that this identity
is satisfied so well for the case of non-spinning black holes.


\section{Discussion}
\label{sec:discussion}

Our purpose in this paper has been to explore the spins of
black holes in equal-mass binaries in order to verify that the
corotating and non-spinning cases were being modeled correctly, and
also to explore the assumptions being used to identify configurations
that are in circular orbits.  In the process of doing this, we have
discovered that the assumptions made in the first attempt to construct
non-spinning equal-mass binaries (using the quasiequilibrium approach
described in this paper)\cite{cook-pfeiffer-2004a} were not leading to
sufficiently accurate representations of non-spinning binaries.
However, the same quasi-local techniques used to measure the spins
of the black holes can also be used to define a new approach for
constructing non-spinning black-hole binaries, and this approach
has produced excellent results.  A detailed description of the
new results is found in Sec.~\ref{sec:results-non-spinning} and
in App.~\ref{sec:no-spin-sequence}.

\begin{table}[!htbp]
\caption{Parameters of the ISCO configuration for {\em non-spinning}
equal-mass black holes.  Results for the ISCO are given for three
different choices of the lapse boundary condition with circular orbits
defined by the Komar-mass ansatz and for a single lapse boundary
condition with circular orbits defined by the EP method.  For
comparison, the lower part of the table lists results of
Refs.~\cite{cook94e,Damour-etal-2002,Blanchet:2002};
``Conf.\ Imag.''~\cite{cook94e} represents data derived from previous
numerical initial-data sets;
``PN standard''~\cite{Blanchet:2002} represents a post-Newtonian
expansion in the standard form without use of the EOB-technique.}
\begin{ruledtabular}
\begin{tabular}{rc|ccc}
Lapse BC & ISCO type 
         & $m\Omega_0$ & $E_b/m$ & $J/m^2$ \\
\hline
$\frac{\td{(\Lapse\CF)}}{\td{r}}=0$ 
         & Komar & 0.122 & -0.0194 & 0.779 \\
\hline
$\frac{\td{(\Lapse\CF)}}{\td{r}}= \frac{\Lapse\CF}{2r} $ 
         & Komar & 0.122 & -0.0194 & 0.779 \\
         & EP & 0.121 & -0.0193 & 0.780 \\
\hline
$\Lapse\CF=\frac12$ 
         & Komar & 0.124 & -0.0195 & 0.778 \\
\hline
\hline
\multicolumn{2}{@{\hspace*{1cm}}l|}{Conf.~Imag.}
                 & 0.166 & -0.0225 & 0.744 \\
\hline
\multicolumn{2}{@{\hspace*{1cm}}l|}{1PN EOB}
                 & 0.0692 & -0.0144 & 0.866 \\
\multicolumn{2}{@{\hspace*{1cm}}l|}{2PN EOB}
                 & 0.0732 & -0.0150 & 0.852 \\
\multicolumn{2}{@{\hspace*{1cm}}l|}{3PN EOB}
                 & 0.0882 & -0.0167 & 0.820\\
\hline
\multicolumn{2}{@{\hspace*{1cm}}l|}{1PN standard}
                 & 0.5224 & -0.0405 & 0.621 \\
\multicolumn{2}{@{\hspace*{1cm}}l|}{2PN standard}
                 & 0.1371 & -0.0199 & 0.779 \\
\multicolumn{2}{@{\hspace*{1cm}}l|}{3PN standard}
                 & 0.1287 & -0.0193 & 0.786
\end{tabular}
\end{ruledtabular}
\label{tab:ISCO_MSTI}
\end{table}

\begin{figure}
\includegraphics[width=\linewidth]{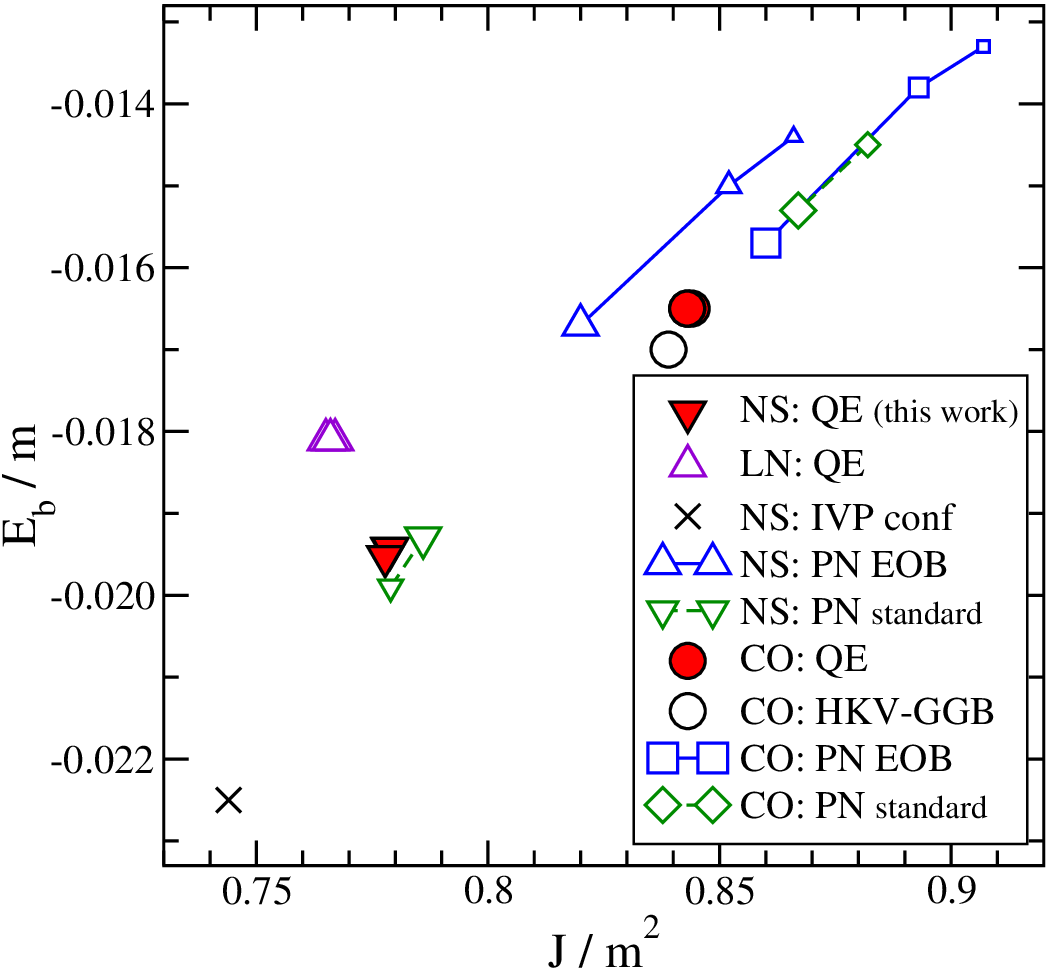}
\caption{ISCO configuration for non-spinning (NS) binary
black holes, computed with three different lapse boundary conditions.
For comparison, the leading-order non-spinning (LN) and corotating
(CO) binary black holes from our earlier
work~\cite{cook-pfeiffer-2004a} are included, as well as results of
Refs.~\cite{cook94e, gourgoulhon-etal-2002b,
Damour-etal-2002,Blanchet:2002}. For post-Newtonian calculations the
size of the symbol indicates the order, the largest symbol being 3PN.
``PN standard''~\cite{Blanchet:2002} represents a PN-expansion in the
standard form without use of the EOB-technique (only 2PN and 3PN are
plotted).}
\label{fig:Eb_J_ISCO}
\end{figure}

\begin{figure}
\includegraphics[width=\linewidth]{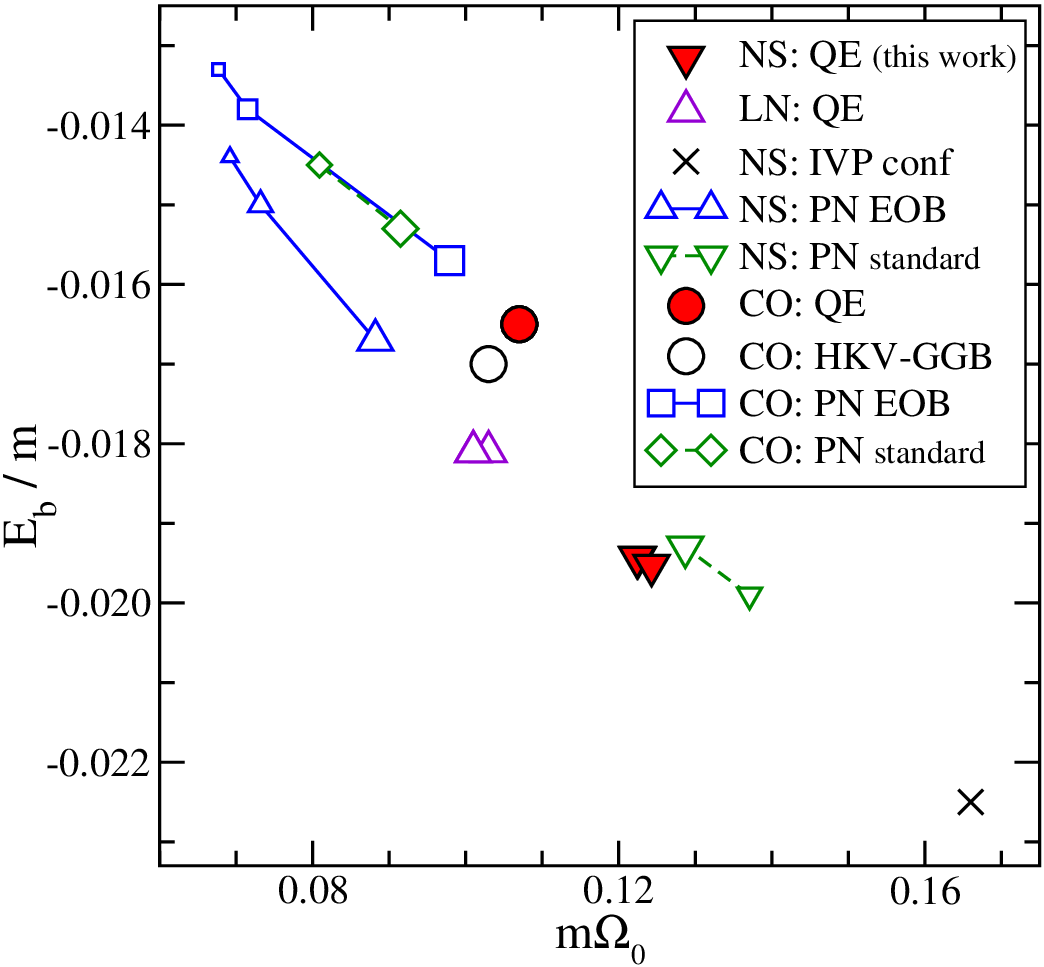}
\caption{ISCO configuration for non-spinning (NS) binary
black holes, computed with three different choices of the lapse
boundary condition.  Symbols as in Fig.~\ref{fig:Eb_J_ISCO}.}
\label{fig:Eb_O_ISCO}
\end{figure}

\begin{figure}
\includegraphics[width=\linewidth]{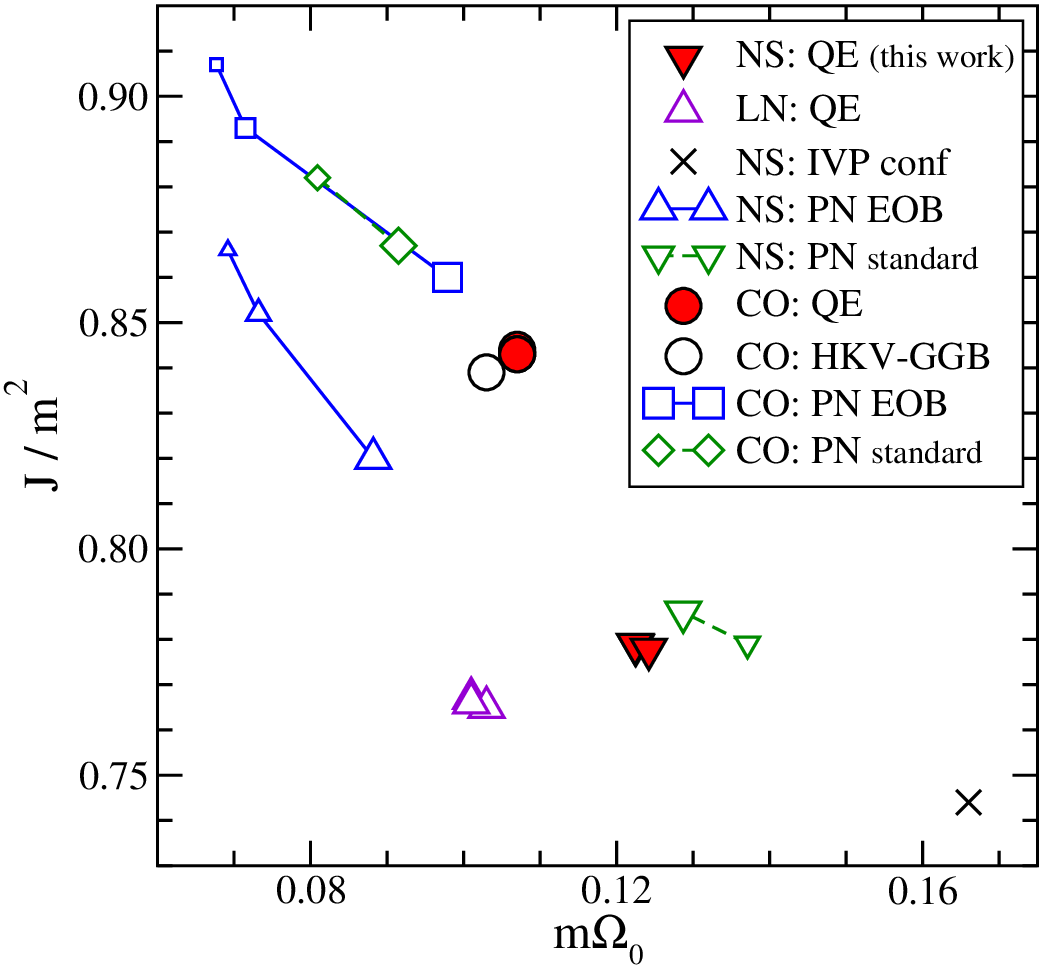}
\caption{ISCO configuration for non-spinning (NS) binary
black holes, computed with three different choices of the lapse
boundary condition.  Symbols as in Fig.~\ref{fig:Eb_J_ISCO}.}
\label{fig:J_O_ISCO}
\end{figure}

For completeness, we also include tables and figures that detail the
ISCO configurations for both corotating and non-spinning cases.
Table~\ref{tab:ISCO_MSTI} displays the new results for ISCO
configurations for non-spinning equal-mass binaries.  These have
changed considerably as is evident by examining these same parameters
plotted in Figs.~\ref{fig:Eb_J_ISCO}--\ref{fig:J_O_ISCO}.  In these
figures, the previous non-spinning results are labeled as {\tt LN: QE}
and the new results as {\tt NS: QE}.  To facilitate direct comparison,
we also include Table~\ref{tab:ISCO_MSCO} which displays the ISCO
result for corotating equal-mass binaries.  In both cases, we include
results obtained using the three choices for the lapse boundary
condition given in Eq.~(\ref{eq:Lapse-BCs}) and with circular orbits
determined using the Komar-mass condition.  Both tables also include a
single ISCO model where the circular orbit is determined using the EP
method.  We include the results only for the lapse boundary condition
in Eq.~(\ref{eq:Lapse-BC-2}) because locating an ISCO model within
the EP approach is very expensive computationally and the results are
not sufficiently different from those obtained via the Komar-mass
condition to warrant the expense.

\begin{table}[!htbp]
\caption{Parameters of the ISCO configuration for {\em corotating}
equal-mass black holes.  Results for the ISCO are given for three
different choices of the lapse boundary condition with circular orbits
defined by the Komar-mass ansatz and for a single lapse boundary
condition with circular orbits defined by the EP method.  Layout as in
Table~\ref{tab:ISCO_MSTI}.}
\begin{ruledtabular}
\begin{tabular}{rc|ccc}
Lapse BC & ISCO type 
         & $m\Omega_0$ & $E_b/m$ & $J/m^2$ \\
\hline
$\frac{\td{(\Lapse\CF)}}{\td{r}}=0$ 
         & Komar & 0.107 & -0.0165 & 0.844 \\
\hline
$\frac{\td{(\Lapse\CF)}}{\td{r}}= \frac{\Lapse\CF}{2r} $ 
         & Komar & 0.107 & -0.0165 & 0.843 \\
         & EP & 0.104 & -0.0163 & 0.845 \\
\hline
$\Lapse\CF=\frac12$ 
         & Komar & 0.107 & -0.0165 & 0.843 \\
\hline
\hline
\multicolumn{2}{@{\hspace*{1cm}}l|}{HKV-GGB}
                 & 0.103 & -0.017 & 0.839 \\
\hline
\multicolumn{2}{@{\hspace*{1cm}}l|}{1PN EOB}
                 & 0.0667 & -0.0133 & 0.907 \\
\multicolumn{2}{@{\hspace*{1cm}}l|}{2PN EOB}
                 & 0.0715 & -0.0138 & 0.893 \\
\multicolumn{2}{@{\hspace*{1cm}}l|}{3PN EOB}
                 & 0.0979 & -0.0157 & 0.860 \\
\hline
\multicolumn{2}{@{\hspace*{1cm}}l|}{1PN standard}
                 & 0.5224 & -0.0405 & 0.621 \\
\multicolumn{2}{@{\hspace*{1cm}}l|}{2PN standard}
                 & 0.0809 & -0.0145 & 0.882 \\
\multicolumn{2}{@{\hspace*{1cm}}l|}{3PN standard}
                 & 0.0915 & -0.0153 & 0.867
\end{tabular}
\end{ruledtabular}
\label{tab:ISCO_MSCO}
\end{table}

We think it worth noting that the value in considering the ISCO
parameters is questionable.  The ISCO, of course, can only be defined
when the effects of radiation reaction are ignored so that the
dynamics is conservative.  The effects of radiation reaction are
certainly not small for equal-mass binaries near ISCO, and so the
physical value of locating the ISCO is limited at best.  However, the
ISCO is a well-defined, unique point in a sequence of (quasi-)circular
orbits and so has value as a point of comparison between various
methods.  It also has value in that it marks the regime where the
quasiequilibrium approximations have clearly broken down.

In addition to yielding improved initial data for non-spinning
equal-mass black-hole binaries, our investigations have shown that the
application of quasi-local methods for determining the spins of black
holes in binary systems yield remarkably good results.  For the case
of corotating binaries, we have shown that the quasi-local spin
measured for the individual black holes is in excellent agreement with
the theoretical expectation for the spin.  However, a nagging question
remains with regard to the computation of quasi-local spins.  This
question is associated with the meaning of ``approximate solutions''
of Killing's equation when no true symmetry exists.  This is an issue
we hope to examine further.  We note that during the preparation of
this manuscript, Schnetter et~al.\cite{Schnetter-etal-2006} have also
raised this issue.

Concerning the theoretical expectation for the spin, we note that the
leading order result, obtained by assuming that a corotating black
hole rotates with an angular velocity equal to the orbital angular
velocity of the binary, is not adequate.  Using Alvi's\cite{Alvi-2000}
leading-order correction to the rate at which a tidal distortion
produced by an orbiting star travels around a black hole, we have
deduced an improved prediction for the spin of a corotating black
hole.  It would be useful if a higher order calculation of Alvi's
tidal distortion rotation rate were available.  The effect of the
corrected rotation rate on post-Newtonian models of corotating
binaries should also be considered, especially when higher order
spin-orbit and spin-spin interaction terms become available.

Finally, we note that there is remarkably good agreement between
sequences of circular orbits defined via the Komar-mass ansatz and by
the EP method.  It is natural to ask which method should be used to
locate configurations in circular orbit.  As described in
Sec.~\ref{sec:effect-potent-meth}, we find that the thermodynamic
identity (\ref{eq:thermo_ident0}) of Friedman
et~al.\cite{friedman-etal-2002} that applies to corotating binaries in
circular orbits is satisfied much more closely by models defined by
the EP method than those that satisfy the Komar-mass ansatz.  While
the EP method yields deviations in the one free quantity ($M_{\rm irr}$) 
that are two orders of magnitude smaller than the Komar-mass condition,
the deviations produced by either approach are very small.  From
a practical perspective, applying the EP method is quite expensive
computationally since it requires finding the minima of a ``function''
that is itself defined by root-finding methods.  For most applications,
it is hard to justify this additional cost.

Perhaps more important than the issue of which approach is better is
the fact that both methods agree so well also for the case of
non-spinning binaries {\em and} that both methods are in good
agreement with the thermodynamic identity of
Eq.~(\ref{eq:thermo_ident0}).  This is surprising because the
identity was intended to apply only in the case of corotating
black-holes binaries.  If the agreement only held at large 
separation, it would be reasonable to assume that the identity
held because strong-field effects were simply small in this
regime.  However, the agreement holds at all separations which
suggests that there may be something more fundamental at work.

\acknowledgments We would like to thank Bernard Whiting and Clifford
Will for useful discussions.  This work was supported in part by NSF
grant PHY-0244906 to the California Institute of Technology.
H.P. gratefully acknowledges support through a Sherman Fairchild Prize
fellowship.  Computations were performed on the Wake Forest University
DEAC Cluster.

\appendix
\section{Solving Killing's equation}
\label{sec:solv-kill-equat}

A straightforward method for finding Killing vectors was outlined
by Dreyer et~al.\cite{dreyer-etal-2003}.  Here we outline that
method as we have implemented it to locate Killing vectors on
a surface ${\cal S}$ with $S^2$ topology.  In our case, ${\cal S}$
is a coordinate sphere in the flat conformal 3-geometry.  Following
Eq.~(\ref{eq:CBMetric_def}), our metric is denoted $\BMetric_{ij}$
and it is conformally related to $\CBMetric_{ij}$ which is the
metric of a coordinate 2-sphere with radius $r$.  Following
Ref.~\cite{dreyer-etal-2003}, we let $\xi^i$ denote a Killing
vector on $({\cal S},\BMetric_{ij})$ and define a two-form
$L_{ij}\equiv\BCD_i\xi_j$, where $\BCD_i$ is the covariant
derivative compatible with $\BMetric_{ij}$.  Now, the Killing
equation can be written simply as
\begin{equation}
\label{eq:Killing_eqn}
L_{(ij)}=0.
\end{equation}
However, instead of solving the Killing equation directly, Dreyer
et~al.\ propose to solve the {\em Killing transport equations}:
\begin{subequations}\label{eq:KT}
\begin{align}
  \label{eq:KT_L}
  \BCD_i\xi_j &= L_{ij}, \\
  \label{eq:KT_dL}
  \BCD_iL_{jk} &= \BRicci_{kji}{}^\ell\xi_\ell,
\end{align}
\end{subequations}
for the two-form $L_{ij}$ and vector $\xi^i$.  Here,
$\BRicci_{kji}{}^\ell$ is the Riemann tensor associated with
$\BMetric_{ij}$.  Note that Eq.~(\ref{eq:KT_dL}) follows from
Eq.~(\ref{eq:KT_L}) by using various symmetries of Riemann {\em
and assuming that $\xi^i$ is a Killing vector} (i.e. that 
$\BCD_{(i}\xi_{j)}=0$).

The central idea in using Eqs.~(\ref{eq:KT}) to
find Killing vectors is to note that the Killing transport equations
act as a linear map on the set of variables $(\xi_i,L_{ij})$.  In
particular, consider integrating these equations along some path from
point $p$ to point $q$.  If $({\cal S},\BMetric_{ij})$ has solutions
of the Killing equation (\ref{eq:Killing_eqn}), and if
$(\xi_i,L_{ij})|_p$ come from one of these Killing solutions, then
integrating the Killing transport equations along the path will yield
$(\xi_i,L_{ij})|_q$ which comes from the {\em same} Killing vector.

Since our surface ${\cal S}$ is 2-dimensional, it follows that
\begin{align}
  \BRicci_{ijk\ell} &= 
            \mbox{$\frac12$}\BRicciS\,\epsilon_{ij}\epsilon_{k\ell}, \\
  L_{ij} &= L\,\epsilon_{ij},
\end{align}
where $\BRicciS$ is the 2-dimensional Ricci scalar associated with
$\BMetric_{ij}$ and $L$ is an oriented scalar on ${\cal S}$.  For our
particular metric,
\begin{equation}
\td{s}^2 = \CF^4 r^2(\td{\theta}^2 + \sin^2\theta\td{\phi}^2),
\end{equation}
the Killing transport equations can be written in spherical
coordinate components as
\begin{subequations}
\begin{align}
\frac{\partial\xi_\theta}{\partial\theta} &=
         2\xi_\theta\partial_\theta\ln\CF 
	 - \frac{2}{\sin^2\theta}\xi_\phi\partial_\phi\ln\CF, \\
\frac{\partial\xi_\phi}{\partial\theta} &=
         \cot\theta\xi_\phi + L\CF^4r^2\sin\theta
         + 2\xi_\theta\partial_\phi\ln\CF 
	 + 2\xi_\phi\partial_\theta\ln\CF, \\
\frac{\partial L}{\partial\theta} &=
         -\left(\frac{1}{\CF^4r^2} - 2\BCD^i\BCD_i\ln\CF\right)
	 \frac{1}{\sin\theta}\xi_\phi,
\end{align}
\end{subequations}
\begin{subequations}
\begin{align}
\frac{\partial\xi_\theta}{\partial\phi} &=
         \cot\theta\xi_\phi - L\CF^4r^2\sin\theta
         + 2\xi_\theta\partial_\phi\ln\CF 
	 + 2\xi_\phi\partial_\theta\ln\CF, \\
\frac{\partial\xi_\phi}{\partial\phi} &=
         -\sin\theta\cos\theta\xi_\theta
         - 2\sin^2\theta\xi_\theta\partial_\theta\ln\CF 
	 + 2\xi_\phi\partial_\phi\ln\CF, \\
\frac{\partial L}{\partial\phi} &=
         \left(\frac{1}{\CF^4r^2} - 2\BCD^i\BCD_i\ln\CF\right)
	 \sin\theta\xi_\theta.
\end{align}
\end{subequations}

It is particularly convenient to compute these quantities in terms of
a basis set $(\bar\theta,\bar\phi)$ which is orthonormal as defined on
the {\em unit} 2-sphere.  We find then that:
\begin{equation}
\label{eq:ortho_basis}
\xi_{\bar\theta} \equiv \xi_\theta \quad\mbox{and}\quad
\xi_{\bar\phi} \equiv \frac{1}{\sin\theta}\xi_\phi.
\end{equation}
The Killing transport equations then take the form
\begin{subequations}
\begin{align}
\label{eq:KT_ott}
\frac{\partial\xi_{\bar\theta}}{\partial\theta} &=
	 2\xi_{\bar\theta}(\SGrad\ln\CF)_{\bar\theta}
	 - 2\xi_{\bar\phi}(\SGrad\ln\CF)_{\bar\phi}, \\
\label{eq:KT_opt}
\frac{\partial\xi_{\bar\phi}}{\partial\theta} &=
         L\CF^4r^2 + 2\xi_{\bar\theta}(\SGrad\ln\CF)_{\bar\phi} 
	 + 2\xi_{\bar\phi}(\SGrad\ln\CF)_{\bar\theta}, \\
\label{eq:KT_olt}
\frac{\partial L}{\partial\theta} &=
         -\frac{\left(1 - 2\SLap\ln\CF\right)}{\CF^4r^2}\xi_{\bar\phi},
\end{align}
\end{subequations}
\begin{subequations}\label{eq:KT_otp_opp_olp}
\begin{align}
\label{eq:KT_otp}
\frac{\partial\xi_{\bar\theta}}{\partial\phi} &=
         \cos\theta\xi_{\bar\phi} - L\CF^4r^2\sin\theta \\
	 & \mbox{\hspace{0.25in}}
	 + 2\sin\theta\xi_{\bar\theta}(\SGrad\ln\CF)_{\bar\phi}
	 + 2\sin\theta\xi_{\bar\phi}(\SGrad\ln\CF)_{\bar\theta}, \nonumber \\
\label{eq:KT_opp}
\frac{\partial\xi_{\bar\phi}}{\partial\phi} &=
         -\cos\theta\xi_{\bar\theta}
         - 2\sin\theta\xi_{\bar\theta}(\SGrad\ln\CF)_{\bar\theta}
         + 2\sin\theta\xi_{\bar\phi}(\SGrad\ln\CF)_{\bar\phi}, \\
\label{eq:KT_olp}
\frac{\partial L}{\partial\phi} &=
         \frac{\left(1 - 2\SLap\ln\CF\right)}{\CF^4r^2}
	 \sin\theta\xi_{\bar\theta},
\end{align}
\end{subequations}
where $\SGrad$ and $\SLap$ are the usual gradient and Laplacian
operators defined for the unit 2-sphere.

Let ${\mathbf V}$ denote the vector of quantities
\begin{equation}
  {\mathbf V} = \left(\begin{array}{c}
                  \xi_{\bar\theta} \\
		  \xi_{\bar\phi} \\
		  L
		\end{array}\right).
\end{equation}
Using fourth-order Runge-Kutta, we integrate
Eqs.~(\ref{eq:KT_otp_opp_olp}) around a closed path from
$\phi=0$ to $\phi=2\pi$ along the equator $\theta=\frac{\pi}{2}$ starting with
three different values for ${\mathbf V}|_{(\phi=0)}$: $(1,0,0)$,
$(0,1,0)$, and $(0,0,1)$.  The three resulting vectors, ${\mathbf
V}|_{(\phi=2\pi)}$ can be used to construct a matrix ${\mathbf M}$
that represents the action of the linear map of the Killing transport
equations on any vector:
\begin{equation}
  \label{eq:Killing_map}
  {\mathbf V}|_{(\phi=2\pi)} = {\mathbf M}\cdot{\mathbf V}|_{(\phi=0)}.
\end{equation}
If ${\mathbf V}|_{(\phi=0)}$ is derived from a Killing vector, then
Eq.~(\ref{eq:Killing_map}) will yield ${\mathbf V}|_{(\phi=2\pi)} =
{\mathbf V}|_{(\phi=0)}$.  So, if $({\cal S},\BMetric_{ij})$ possesses
a global solution of the Killing equation, it will be associated with
a {\em unit eigenvalue} of ${\mathbf M}$ and the associated
eigenvector will be derived from that Killing vector.  In general,
$({\cal S},\BMetric_{ij})$ will not possess a Killing vector and
${\mathbf M}$ will have no unit eigenvalues.  One could proceed to
construct an ``approximate Killing vector'' by using an eigenvalue
that is sufficiently close to unity.

However, for the corotating and non-spinning black-hole binary initial
data we consider in this paper, the conformal factor possesses a
reflection symmetry through the plane of the orbit.  We thus find that
the eigenvector $(0,1,0)$ always has unit eigenvalue when we integrate
along the equator.\footnote{This is most easily seen by writing the
Killing transport equations in terms of an orthonormal basis set
$(\hat\theta,\hat\phi)$ defined with respect to the full metric
$\BMetric_{ij}$ so that $\xi_{\hat\theta}\equiv\frac1{\CF^2
r}\xi_\theta$ and $\xi_{\hat\phi}\equiv\frac1{\CF^2
r\sin\theta}\xi_\phi$.}  Unfortunately, this vector does not
necessarily represent a global Killing vector on ${\cal S}$.

Our pseudo-spectral code for solving the constraint equations
represents quantities on the excision boundaries in terms of spherical
harmonic decompositions.  To proceed we need to populate the
collocation points (grid points) of the spectral grid with values for
${\mathbf V}$ obtained by propagating the eigenvector $(0,1,0)$ at
$(\theta=\frac{\pi}2,\phi=0)$ to all of these points via the Killing
transport equations.  If our starting vector were constructed from a
global Killing vector, then the path we take to each of these points
would not matter.  However, since we do not in general have a true
global Killing vector, we must specify how we populate the grid points.

First, we integrate Eqs.~(\ref{eq:KT_otp_opp_olp}) to
populate values along the equator ($\theta=\frac{\pi}2$) wherever they
are needed, integrating from $\phi=0$ to $\phi=2\pi$.  For each line
of colatitude containing collocation points, we integrate
Eqs.~(\ref{eq:KT_ott}--\ref{eq:KT_olt}) starting with the known values
at the equator.  All integrations are performed using fourth-order
Runge-Kutta.  With values for $\xi_{\hat\theta}$ and $\xi_{\hat\phi}$
at all collocation points we can construct a vector spherical harmonic
representation of our solution over the entire surface ${\cal S}$.
All that remains is to normalize the Killing vector and check to see
if we have a true Killing vector.

A rotational Killing vector (as we are trying to construct) is
normalized so that its affine length is $2\pi$.  Consider
a path parameterized by $t$ and defined by
\begin{subequations}
\begin{align}
\frac{\partial\theta}{\partial t} &= \frac{1}{\CF^4r^2}\xi_{\bar\theta}, \\
\frac{\partial\phi}{\partial t} &= \frac{1}{\CF^4r^2\sin\theta}\xi_{\bar\phi}.
\end{align}
\end{subequations}
We integrate from some starting location $(\theta,\phi)|_0$ with $t=0$
until the path closes at the starting point.  If ${\mathbf\xi}$ is not
properly normalized, then the final value of $t_f$ will not be $2\pi$.
However, using $t_f$ we can rescale ${\vec\xi}$ so that it does have
an affine length of $2\pi$.  In practice, we normalize the solution by
integrating along the equator, then check that the solution is
correctly normalized to within truncation error for several different
integral paths starting at $\phi=0$ and various initial values for
$\theta$.

There are many possible ways of determining whether or not our
solution is a global solution of the Killing equation.  For
simplicity, we check the following set of scalar conditions that must
be satisfied everywhere if our solution is a true Killing vector of
$({\cal S},\BMetric_{ij})$:
\begin{subequations}
\begin{align}
\label{eq:KVdiag_trace0}
\xi^i\BCD_iL &= 0, \\
\label{eq:KVdiag_div0}
\BCD_i\xi^i &= 0, \\
\label{eq:KVdiag_curl0}
\epsilon^{ij}\BCD_i\xi_j &= 2L.
\end{align}
\end{subequations}
In terms of our basis defined by Eq.~(\ref{eq:ortho_basis}), these become
\begin{subequations}
\begin{align}
\label{eq:KVdiag_trace}
\frac{1}{\CF^4r^2}\left(\xi_{\bar\theta}(\SGrad L)_{\bar\theta} 
        + \xi_{\bar\phi}(\SGrad L)_{\bar\phi}\right) &= 0, \\
\label{eq:KVdiag_div}
\frac{1}{\CF^4r^2}\SDiv\vec\xi &= 0, \\
\label{eq:KVdiag_curl}
\frac{1}{\CF^4r^2}\SVort\vec\xi &= 2 L,
\end{align}
\end{subequations}
and $\SDiv$ and $\SVort$ are the usual divergence and twist
operators defined for the unit 2-sphere.  We find that these
identities are never satisfied for our solutions although
the residuals seem ``small'' and decrease in size as the
separation between the black holes increases.  Because we have
not yet found any way to meaningfully normalize the values of
the residuals, we do no bother to report their values here.

We do note that, because we represent $\vec\xi$ in terms of vector
spherical harmonics, it is possible to filter the coefficients in the
vector spherical harmonic expansion to ensure that it is
divergenceless.  That is, we can, {\em a posteriori} guarantee that
Eq.~(\ref{eq:KVdiag_div}) is satisfied.  However, this does not
change the other diagnostics.  All of the results presented 
in this paper that depend on solutions of the Killing equation
have been filtered in this way.  We note that this filtering
has a negligible effect on all measured quantities.

We conclude this Appendix by noting that our solutions of the Killing
equation are ``approximate Killing vectors'' in some sense that is not
at all well defined.  In all cases, we find that our approximate
Killing vector is very similar to the corresponding solution on the unit
2-sphere.  These are conformal Killing vectors of $({\cal
S},\BMetric_{ij})$ as written down in
Eqs.~(\ref{eq:CKV_x}--\ref{eq:CKV_z}).  We leave it to future work to
more rigorously define the meaning of these ``approximate Killing
vectors.''  

Finally, we note that during the preparation of this manuscript,
Schnetter et~al.\cite{Schnetter-etal-2006} have shown that when a
divergence free approximate Killing vector is used to define the
quasi-local angular momentum, as in Eq.~(\ref{eq:QL_spin}), then it is
gauge invariant.  Therefore, they suggest that such a divergence-free
approximate Killing vector ``can be viewed as an {\em ersatz} axial
symmetry vector even in the absence of axisymmetry.''

\section{Corotating sequence}
\label{sec:corotating-sequence}
In this Appendix, we list the numerical results for corotating
equal-mass black holes in a quasi-circular orbit as defined by
the Komar ansatz.  We have assumed conformal flatness, maximal slicing,
and used Eq.~(\ref{eq:Lapse-BC-2}) for the lapse boundary condition
on both excision surfaces.  The data is scaled relative to the
total and reduced masses ($m$ and $\mu$) defined with respect to
the irreducible mass of the apparent horizons $M_{\rm irr}$.

In Table~\ref{tab:corot_D_MS}, $d$ and $\Omega_0$ are, respectively,
the separation of the centers of the excised regions and the orbital
angular velocity measured in ``coordinate units''.  These values for
$d$ and $\Omega$, together with the coordinate radius of the excision
sphere $r=0.7857981371$, provide all parameters necessary to reproduce
the data in Table~\ref{tab:corot_D_MS}.  The remaining quantities are
dimensionless.  $m\Omega_0$ is the orbital angular velocity of the
binary system as measured at infinity.  $E_b/\mu$ is the dimensionless
binding energy of the system with $E_b$ defined as $E_b\equiv
E_{\mbox{\tiny ADM}}\!\!\!-m$.  $J/\mu{m}$ is the dimensionless total
ADM angular momentum of the binary system as measured at infinity.
$\ell/m$ is the dimensionless proper separation between the two
excision surfaces as measured on the initial-data slice.\footnote{
Note that the value of $\ell$ listed in Tables IV and V of
Ref.\cite{cook-pfeiffer-2004a} were in error.  The corrected values
for the corotating case, and correct values for the ``true'' non-spinning
case, are given in the following tables.}
Finally, $S_z/M_{\rm irr}^2$ and $S_K/M_{\rm irr}^2$ are two measures
of the dimensionless spin of one of the black holes.  $S_z$ is defined
using the flat space Killing vector and $S_K$ is defined using the
approximate Killing vector.

\pagebreak
\vspace*{0.8in}

\begin{widetext}
\onecolumngrid
\begin{table}[]
\caption{Sequence of corotating equal-mass black holes on circular orbits
satisfying the Komar ansatz.  The ISCO is at separation $d\sim8.28$.}
\begin{ruledtabular}
\begin{tabular}{ll|llllll}
\multicolumn{1}{c}{$d$} & \multicolumn{1}{c|}{$\Omega_0$} 
& \multicolumn{1}{c}{$m\Omega_0$} 
& \multicolumn{1}{c}{$E_b/\mu$} 
& \multicolumn{1}{c}{$J/\mu{m}$} 
& \multicolumn{1}{c}{$\ell/m$}
& \multicolumn{1}{c}{$S_z/M_{\rm irr}^2$}
& \multicolumn{1}{c}{$S_K/M_{\rm irr}^2$} \\
\hline
40   & 0.0052966 & 0.01090 & -0.0233183 & 4.9120 & 22.87 & 0.02148 & 0.02150 \\
35   & 0.0064212 & 0.01327 & -0.0263265 & 4.6620 & 20.23 & 0.02609 & 0.02612 \\
30   & 0.0080083 & 0.01665 & -0.0301930 & 4.4019 & 17.58 & 0.03264 & 0.03267 \\
29   & 0.0084050 & 0.01750 & -0.0311008 & 4.3488 & 17.04 & 0.03428 & 0.03431 \\
28   & 0.0088354 & 0.01842 & -0.0320622 & 4.2953 & 16.51 & 0.03606 & 0.03610 \\
27   & 0.0093039 & 0.01943 & -0.0330816 & 4.2414 & 15.97 & 0.03800 & 0.03804 \\
26   & 0.0098151 & 0.02054 & -0.0341642 & 4.1872 & 15.43 & 0.04012 & 0.04017 \\
25   & 0.010375  & 0.02175 & -0.0353153 & 4.1328 & 14.89 & 0.04245 & 0.04250 \\
24   & 0.010990  & 0.02309 & -0.0365409 & 4.0781 & 14.35 & 0.04501 & 0.04507 \\
23   & 0.011667  & 0.02457 & -0.0378476 & 4.0233 & 13.81 & 0.04785 & 0.04791 \\
22   & 0.012418  & 0.02622 & -0.0392423 & 3.9684 & 13.26 & 0.05099 & 0.05106 \\
21   & 0.013252  & 0.02806 & -0.0407325 & 3.9135 & 12.72 & 0.05449 & 0.05457 \\
20   & 0.014183  & 0.03012 & -0.0423262 & 3.8587 & 12.17 & 0.05841 & 0.05850 \\
19   & 0.015227  & 0.03245 & -0.0440313 & 3.8043 & 11.62 & 0.06282 & 0.06293 \\
18   & 0.016406  & 0.03511 & -0.0458555 & 3.7503 & 11.06 & 0.06782 & 0.06794 \\
17   & 0.017744  & 0.03814 & -0.0478056 & 3.6971 & 10.50 & 0.07352 & 0.07366 \\
16   & 0.019273  & 0.04164 & -0.0498857 & 3.6450 & 9.943 & 0.08007 & 0.08024 \\
15   & 0.021031  & 0.04571 & -0.0520956 & 3.5944 & 9.377 & 0.08765 & 0.08785 \\
14.5 & 0.022012  & 0.04800 & -0.0532470 & 3.5699 & 9.092 & 0.09190 & 0.09212 \\
14   & 0.023070  & 0.05049 & -0.0544267 & 3.5460 & 8.807 & 0.09650 & 0.09674 \\
13.5 & 0.024215  & 0.05320 & -0.0556310 & 3.5229 & 8.520 & 0.1015  & 0.1018 \\
13   & 0.025456  & 0.05617 & -0.0568549 & 3.5006 & 8.231 & 0.1070  & 0.1073 \\
12.5 & 0.026804  & 0.05942 & -0.0580912 & 3.4793 & 7.941 & 0.1129  & 0.1132 \\
12   & 0.028273  & 0.06300 & -0.0593298 & 3.4592 & 7.649 & 0.1195  & 0.1198 \\
11.5 & 0.029877  & 0.06696 & -0.0605565 & 3.4405 & 7.356 & 0.1266  & 0.1271 \\
11   & 0.031634  & 0.07135 & -0.0617523 & 3.4234 & 7.060 & 0.1346  & 0.1350 \\
10.5 & 0.033564  & 0.07625 & -0.0628909 & 3.4082 & 6.762 & 0.1434  & 0.1439 \\
10   & 0.035689  & 0.08174 & -0.0639364 & 3.3952 & 6.462 & 0.1532  & 0.1538 \\
9.5  & 0.038037  & 0.08792 & -0.0648401 & 3.3848 & 6.159 & 0.1642  & 0.1649 \\
9    & 0.040638  & 0.09493 & -0.0655348 & 3.3775 & 5.853 & 0.1765  & 0.1773 \\
8.9  & 0.041191  & 0.09644 & -0.0656410 & 3.3765 & 5.792 & 0.1792  & 0.1800 \\
8.8  & 0.041757  & 0.09800 & -0.0657345 & 3.3756 & 5.731 & 0.1819  & 0.1828 \\
8.7  & 0.042334  & 0.09960 & -0.0658142 & 3.3749 & 5.669 & 0.1847  & 0.1856 \\
8.6  & 0.042923  & 0.1012  & -0.0658791 & 3.3744 & 5.607 & 0.1876  & 0.1885 \\ 
8.5  & 0.043526  & 0.1029  & -0.0659282 & 3.3740 & 5.545 & 0.1905  & 0.1915 \\
8.4  & 0.044141  & 0.1047  & -0.0659603 & 3.3738 & 5.483 & 0.1936  & 0.1945 \\
8.35 & 0.044453  & 0.1055  & -0.0659695 & 3.3737 & 5.451 & 0.1951  & 0.1961 \\
8.3  & 0.044769  & 0.1064  & -0.0659741 & 3.3738 & 5.420 & 0.1967  & 0.1977 \\
8.25 & 0.045089  & 0.1074  & -0.0659738 & 3.3738 & 5.389 & 0.1982  & 0.1993 \\
8.2  & 0.045412  & 0.1083  & -0.0659685 & 3.3739 & 5.358 & 0.1998  & 0.2009 \\
8.1  & 0.046068  & 0.1102  & -0.0659419 & 3.3743 & 5.295 & 0.2031  & 0.2042 \\
8    & 0.046738  & 0.1121  & -0.0658929 & 3.3748 & 5.232 & 0.2065  & 0.2076
\end{tabular}
\end{ruledtabular}
\label{tab:corot_D_MS}
\end{table}
\end{widetext}

\section{Non-spinning sequence}
\label{sec:no-spin-sequence}
In this Appendix, we list the numerical results for non-spinning
equal-mass black holes in a quasi-circular orbit as defined by
the Komar ansatz.  We have assumed conformal flatness, maximal slicing,
and used Eq.~(\ref{eq:Lapse-BC-2}) for the lapse boundary condition
on both excision surfaces.  The data is scaled relative to the
total and reduced masses ($m$ and $\mu$) defined with respect to
the irreducible mass of the apparent horizons $M_{\rm irr}$.

In Table~\ref{tab:irrot_D_MS}, $d$ and $\Omega_0$ are, respectively,
the separation of the centers of the excised regions and the orbital
angular velocity measured in ``coordinate units''.
$f_r=\Omega_r/\Omega_0$ is the rotation fraction necessary to obtain a
non-spinning black hole.  These values for $d$, $\Omega$, and $f_r$,
together with the coordinate radius of the excision sphere
$r=0.7857981371$, provide all parameters necessary to reproduce the
data in Table~\ref{tab:irrot_D_MS}.  The remaining quantities are
dimensionless.  $m\Omega_0$ is the orbital angular velocity of the
binary system as measured at infinity.  $E_b/\mu$ is the dimensionless
binding energy of the system with $E_b$ defined as $E_b\equiv
E_{\mbox{\tiny ADM}}\!\!\!-m$.  $J/\mu{m}$ is the dimensionless total
ADM angular momentum of the binary system as measured at infinity.
$\ell/m$ is the dimensionless proper separation between the two
excision surfaces as measured on the initial-data slice.  $S_z/M_{\rm
irr}^2$ is the the dimensionless spin of one of the black holes where
$S_z$ is defined using the flat space Killing vector.  The spin
defined using the approximate Killing vector is zero by definition.

\begin{widetext}
\onecolumngrid
\begin{table}[]
\caption{Sequence of non-spinning equal-mass black holes on circular
orbits satisfying the Komar ansatz.  The ISCO is at separation
$d\sim7.55$.}
\begin{ruledtabular}
\begin{tabular}{lll|lllll}
\multicolumn{1}{c}{$d$} & \multicolumn{1}{c}{$\Omega_0$} 
& \multicolumn{1}{c|}{$f_r$} 
& \multicolumn{1}{c}{$m\Omega_0$} 
& \multicolumn{1}{c}{$E_b/\mu$} 
& \multicolumn{1}{c}{$J/\mu{m}$} 
& \multicolumn{1}{c}{$\ell/m$}
& \multicolumn{1}{c}{$S_z/M_{\rm irr}^2$}\\
\hline
40   & 0.0052973 & 0.9551 & 0.01090 & -0.0235329 & 4.8717 & 22.87 & -0.00004206 \\
35   & 0.0064225 & 0.9485 & 0.01327 & -0.0266396 & 4.6136 & 20.24 & -0.00005974 \\
30   & 0.0080108 & 0.9396 & 0.01665 & -0.0306751 & 4.3423 & 17.58 & -0.00008963 \\
29   & 0.0084079 & 0.9374 & 0.01750 & -0.0316304 & 4.2864 & 17.05 & -0.00009813 \\
28   & 0.0088388 & 0.9351 & 0.01843 & -0.0326458 & 4.2298 & 16.51 & -0.0001077  \\
27   & 0.0093079 & 0.9326 & 0.01943 & -0.0337268 & 4.1727 & 15.98 & -0.0001187  \\
26   & 0.0098198 & 0.9299 & 0.02054 & -0.0348800 & 4.1150 & 15.44 & -0.0001312  \\
25   & 0.010380  & 0.9270 & 0.02175 & -0.0361122 & 4.0568 & 14.90 & -0.0001458  \\
24   & 0.010996  & 0.9238 & 0.02309 & -0.0374317 & 3.9979 & 14.36 & -0.0001627  \\
23   & 0.011676  & 0.9203 & 0.02457 & -0.0388473 & 3.9386 & 13.82 & -0.0001825  \\
22   & 0.012428  & 0.9165 & 0.02622 & -0.0403693 & 3.8786 & 13.27 & -0.0002058  \\
21   & 0.013264  & 0.9123 & 0.02806 & -0.0420092 & 3.8182 & 12.73 & -0.0002336  \\
20   & 0.014198  & 0.9077 & 0.03013 & -0.0437799 & 3.7574 & 12.18 & -0.0002669  \\
19   & 0.015247  & 0.9025 & 0.03246 & -0.0456957 & 3.6962 & 11.63 & -0.0003072  \\
18   & 0.016431  & 0.8967 & 0.03512 & -0.0477726 & 3.6348 & 11.07 & -0.0003567  \\
17   & 0.017775  & 0.8903 & 0.03816 & -0.0500280 & 3.5732 & 10.52 & -0.0004180  \\
16   & 0.019313  & 0.8829 & 0.04166 & -0.0524802 & 3.5118 & 9.957 & -0.0004952  \\
15   & 0.021085  & 0.8745 & 0.04574 & -0.0551473 & 3.4509 & 9.393 & -0.0005940  \\
14.5 & 0.022075  & 0.8698 & 0.04804 & -0.0565665 & 3.4207 & 9.110 & -0.0006539  \\
14   & 0.023143  & 0.8647 & 0.05053 & -0.0580448 & 3.3907 & 8.825 & -0.0007226  \\
13.5 & 0.024300  & 0.8593 & 0.05326 & -0.0595830 & 3.3612 & 8.539 & -0.0008020  \\
13   & 0.025556  & 0.8533 & 0.05623 & -0.0611807 & 3.3321 & 8.252 & -0.0008942  \\
12.5 & 0.026923  & 0.8469 & 0.05950 & -0.0628363 & 3.3036 & 7.963 & -0.001002   \\
12   & 0.028415  & 0.8399 & 0.06310 & -0.0645457 & 3.2758 & 7.673 & -0.001129   \\
11.5 & 0.030047  & 0.8321 & 0.06709 & -0.0663017 & 3.2490 & 7.381 & -0.001279   \\
11   & 0.031840  & 0.8236 & 0.07151 & -0.0680922 & 3.2233 & 7.087 & -0.001460   \\
10.5 & 0.033815  & 0.8142 & 0.07645 & -0.0698984 & 3.1991 & 6.791 & -0.001678   \\
10   & 0.035998  & 0.8036 & 0.08200 & -0.0716907 & 3.1766 & 6.493 & -0.001946   \\
9.5  & 0.038420  & 0.7918 & 0.08826 & -0.0734242 & 3.1565 & 6.193 & -0.002277   \\
9    & 0.041118  & 0.7785 & 0.09538 & -0.0750303 & 3.1393 & 5.891 & -0.002693   \\
8.5  & 0.044133  & 0.7632 & 0.1035  & -0.0764029 & 3.1258 & 5.585 & -0.003226   \\
8    & 0.047516  & 0.7457 & 0.1129  & -0.0773785 & 3.1172 & 5.277 & -0.003920   \\
7.9  & 0.048241  & 0.7419 & 0.1150  & -0.0775054 & 3.1162 & 5.215 & -0.004084   \\
7.8  & 0.048984  & 0.7379 & 0.1171  & -0.0776039 & 3.1154 & 5.153 & -0.004257   \\
7.7  & 0.049745  & 0.7338 & 0.1193  & -0.0776711 & 3.1149 & 5.090 & -0.004442   \\
7.65 & 0.050132  & 0.7318 & 0.1204  & -0.0776920 & 3.1148 & 5.059 & -0.004538   \\
7.6  & 0.050524  & 0.7296 & 0.1216  & -0.0777037 & 3.1148 & 5.028 & -0.004637   \\
7.55 & 0.050921  & 0.7275 & 0.1227  & -0.0777058 & 3.1148 & 4.996 & -0.004740   \\
7.5  & 0.051323  & 0.7253 & 0.1239  & -0.0776980 & 3.1149 & 4.965 & -0.004846   \\
7.4  & 0.052140  & 0.7208 & 0.1263  & -0.0776501 & 3.1154 & 4.902 & -0.005067   \\
7.3  & 0.052978  & 0.7161 & 0.1288  & -0.0775557 & 3.1163 & 4.839 & -0.005304
\end{tabular}
\end{ruledtabular}
\label{tab:irrot_D_MS}
\end{table}
\end{widetext}

\end{document}